\documentclass[preprint,12pt]{elsarticle}

\usepackage{graphicx}
\usepackage{amssymb}
\usepackage{amsthm}
\usepackage{amsmath}
\usepackage{bm}
\usepackage{fontawesome}
\usepackage{xcolor}
\usepackage{lineno}
\usepackage[unicode]{hyperref}
\usepackage[utf8]{inputenc}
\usepackage{cleveref}
\usepackage{adjustbox}
\usepackage{xcolor-material}
\newcommand\myshade{85}
\colorlet{mylinkcolor}{MaterialRed}
\colorlet{mycitecolor}{MaterialBlue}
\colorlet{myurlcolor}{MaterialGreen}

\hypersetup{
	linkcolor  = mylinkcolor!\myshade!black,
	citecolor  = mycitecolor!\myshade!black,
	urlcolor   = myurlcolor!\myshade!black,
	colorlinks = true,
	linktoc=all
}
\newcommand*{\SavedEqref}{}
\let\SavedEqref\eqref
\renewcommand*{\eqref}[1]{%
	\begingroup
	\hypersetup{
		linkcolor=myurlcolor,
		linkbordercolor=myurlcolor,
	}%
	\SavedEqref{#1}%
	\endgroup
}

\DeclareMathOperator*{\argmax}{arg\,max}

\journal{arXiv}

\begin{document}
\newcommand{\mathbfrm}[1]{\bm{\mathrm{#1}}}
\newcommand{\intdiff}[1]{\ \mathrm{d}#1}
\newcommand{\norm}[1]{\vert\vert#1\vert\vert}
\newcommand{\revised}[1]{\textcolor{black}{#1}}
\newcommand{\myvector}[1]{\mathbfrm{#1}}
\newcommand{\mymatrix}[1]{\mathbfrm{#1}}
\newcommand{\myscalar}[1]{#1}
\newcommand{\mytensor}[1]{\mathcal{#1}}
\begin{frontmatter}


	\title{Localising the Smallest Stiffness and its Direction of a \revised{Homogeneous} Structure by Spectral and Optimisation \revised{Approaches}.}

	\author[1]{Petr Henyš}
	\author[1]{Danas Sutula}
	\author[3]{Jiří Kopal}
	\author[2]{Michal Kuchař}
	\cortext[cor1]{Corresponding author}
	\ead{michal.kuchar@lfhk.cz}
	\author[1]{Lukáš Čapek}
	\address[1]{Department of Technologies and Structures, Textile Faculty, Technical University of Liberec.Studentská 1402/2, 461 17 Liberec, Czech Republic}
	\address[2]{Department of Anatomy, Faculty of Medicine in Hradec Králové, Charles University, Šimkova 870, 500 03 Hradec Králové, Czech Republic}
	\address[3]{Faculty of Mechatronics, Informatics and Interdisciplinary Studies, Technical University of Liberec, Studentská 1402/2, 461 17 Liberec, Czech Republic}
	\begin{abstract}
		Structural stiffness plays an important role in engineering design. The analysis of stiffness requires precise experiments and computational models that can be difficult or time--consuming to procure. \revised{A novel relation between modal and static stiffness based on modal decomposition is introduced in this study. This relation allows analysing the smallest structural stiffness and its direction. Further, it is shown that the smallest stiffness can be found using an optimisation algorithm that is based on the maximisation of structural compliance. Both approaches are compared on several computational examples leading to similar results in terms of smallest stiffness and its direction.} The proposed approaches serve as quantitative/qualitative tools for the analyses of structural stiffness, \revised{particularly in structural health monitoring.}
	\end{abstract}
	\begin{keyword}
		Structure design, \revised{compliance maximisation}, finite element method, modal analysis, reduced--order model
	\end{keyword}
\end{frontmatter}
\revised{
	\section*{Highlights}
	\begin{itemize}
		\item Novel relation between the modal and the static stiffness of a structure.
		\item Identifiability of smallest structural stiffness based on modal properties.
		\item Alternative method for localising the smallest structural stiffness based on structural compliance maximisation.
	\end{itemize}}
\section*{Introduction}
Modal properties provide useful information about the dynamic behaviour of a given structure. The most important property is the so-called eigenfrequency and the associated eigenvector. Those quantities provide unique structural information and are fundamental in the design and optimisation of engineering structures \cite{ZANARINI2019817,CHEN20145566,:/content/journals/10.1049/ip-c.1988.0053,JURECZKO2005463,WANG2016123}. \revised{In analogy with shape analysis \cite{ruggeri2010spectral}, the eigenvalues are invariant to a rigid transformation and do not require the boundary conditions for the associated differential operator to be defined \cite{henyvs2017material}. Hence the modal analysis allows the comparison of different objects from both quantitative and qualitative points of view.}

\begin{figure}[!bht]
	\centering
	\includegraphics[width=140mm]{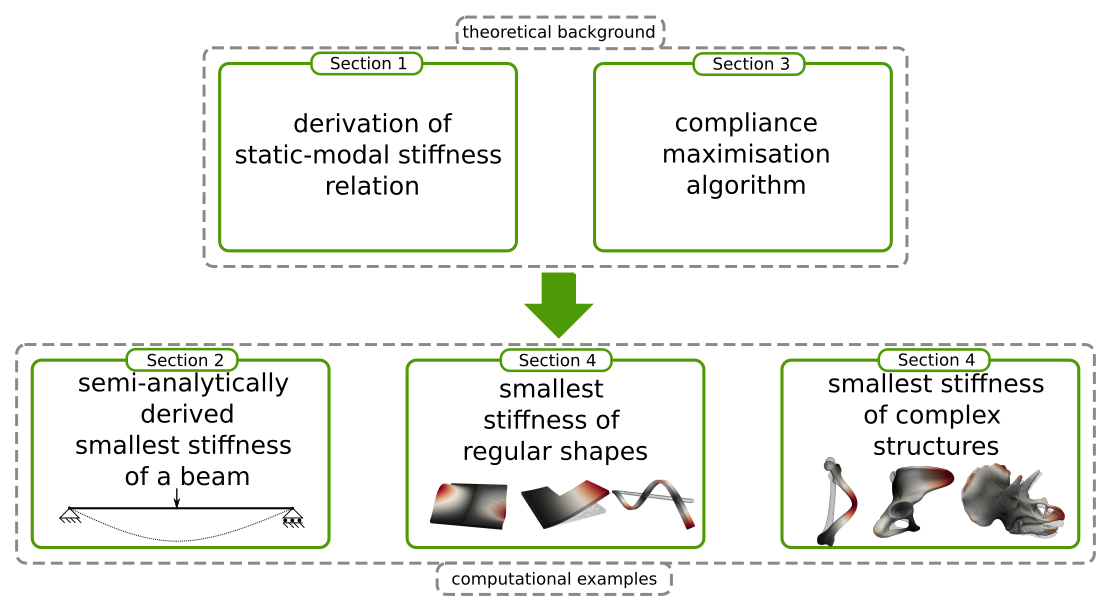}
	\caption{\revised{A study flowchart}}
	\label{fig:studyflow}
\end{figure}

\revised{The modal properties are} closely related to the term stiffness, which can be defined as ``\emph{the capacity of the structure to resist deformation}''. It depends on both the material and geometric properties. The stiffness is one of the key objectives of an engineering design. It is of interest in many engineering applications \cite{bae2019calculation,bagheri2017structural,feng2017identification,jiang2018structural,macleod2018exploring,su2017identification} and is often analysed with help of modal analysis \cite{helsen2010global,zanardo2006stiffness,steiger2012comparison,west2019extraction,chandgude2019investigation}. \revised{The modal analysis provides a modal stiffness, which is not the same as static stiffness; however, since both are defined for the same structure, there is a relation between them. Melnikov et al. \cite{melnikov2017determination} introduced a reconstruction of a stiffness matrix from operational modal data with mass modification. Although the method can reconstruct bending and torsional static stiffness, it strongly depends on the number of mass modification points and modes included, moreover, it is not clear how to choose mass modification rescaling. Helsen et al. \cite{helsen2010global} proposed identification of static stiffness using modal flexibility matrix at zero Hz. This approach is relatively accurate but requires a full modal compliance matrix to be determined. Poland et al. \cite{poland2015estimation,deleener2010extraction} found that torsional and bending stiffness can be estimated from the frequency response matrix estimated on a few degrees of freedom with a good accuracy.}
\revised{The proposed methods are able to accurately estimate the bending and torsional stiffness, but require much additional information such as the number of testing points and their optimal locations. They were specifically developed for the automotive industry for specific experiments, e.g. a car body. From a qualitative point of view, it seems that those methods are unable to naturally detect the smallest stiffness and its location under general conditions.}
\revised{
	This study aims to provide both quantitative and qualitative description of structural stiffness, particularly the smallest stiffness. This can be seen as the main novelty of the study. The workflow of the current study is shown in Figure \ref{fig:studyflow}. Firstly, the relation of modal and static stiffness is derived from truncated spectral decomposition (Section \ref{s:Relation of Static and Modal Stiffness}). Consequently, the hypothesis that the smallest static stiffness and its direction can be found from truncated spectral decomposition is formulated (Section \ref{s:Relation of Static and Modal Stiffness}). In Section \ref{ss:beamstiffnessanalysis}, the hypothesis is demonstrated by analysing a beam stiffness under two types of boundary conditions. As an alternative to spectral analysis, an optimisation algorithm based on strain energy maximisation was developed for finding the smallest stiffness (Section \ref{s:gradientbasedmethodtofindcriticalloadingpoint}). Both approaches are tested and compared on examples of regular and complex shape structures in Section \ref{s:computationalmodels}.}
The following notations are used. Vectors and matrices are indicated by a bold letter. Point quantities are equipped with a lower index. The operator $\norm{\cdot}$ represents $L_{2}$ norm if not stated otherwise.
\section{Relation of Static to Modal Stiffness}
\label{s:Relation of Static and Modal Stiffness}
\begin{figure}[!bht]
	\centering
	\includegraphics[width=90mm]{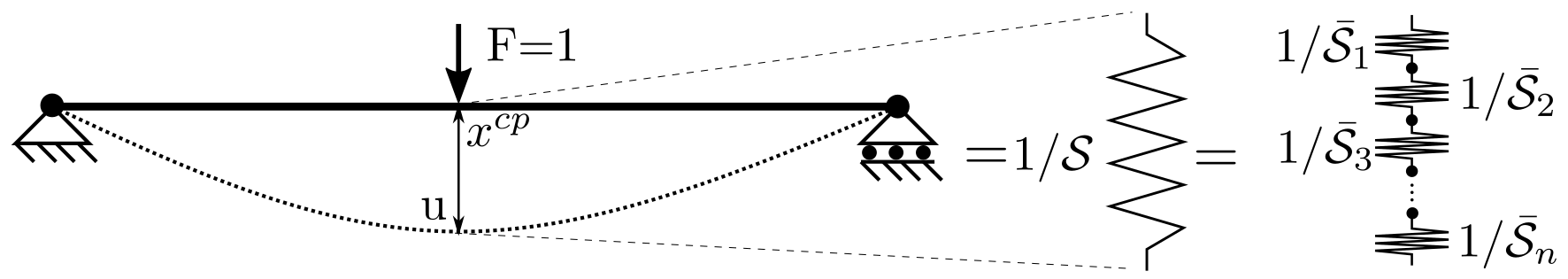}
	\caption{Relation of static $\mytensor{S}$ to modal stiffness $\bar{\mytensor{S}}$ is given by the sum of modal static contributions from each shape mode. The sum is equal to a static stiffness at a given critical point $\myscalar{x}^{cp}$, the details are explained later.}
	\label{fig:staticmodalstiffnessrelation}
\end{figure}

The pointwise smallest static stiffness in a unit load direction $\myvector{t}$ is the smallest possible stiffness magnitude defined over all points on the structure:
\begin{equation}
	\label{eq:minS}
	\norm{\mytensor{S}} = min(\norm{\mytensor{S}_{1}}, \norm{\mytensor{S}_{2}}, \norm{\mytensor{S}_{j}}, \dots, \norm{\mytensor{S}_{N}})
\end{equation}
where $j$ represents $j$'th point and \textit{N} the number of points. In terms of static compliance $\norm{\mytensor{C}}$, \revised{the goal is to find} the point of maximum static displacement magnitude $\norm{\myvector{u}}$ induced with the unit load in direction $\myvector{t}$:
\begin{equation}
	\label{eq:maxU}
	\norm{\mytensor{C}} = max(\norm{\myvector{u}_{1}}, \norm{\myvector{u}_{2}}, \norm{\myvector{u}_{j}},\dots,\norm{\myvector{u}_{N}})
\end{equation}
The point satisfying either \eqref{eq:minS} or \eqref{eq:maxU} is called a critical point $\myvector{x}^{cp}$. This point can be identified intuitively for simple structures such as a cantilever beam where the critical point $\myvector{x}^{cp}$ is located at the free end or for a simply supported beam $\myvector{x}^{cp}$ is midway along the beam as shown in Figure \ref{fig:staticmodalstiffnessrelation}.
\subsection{\revised{Formulation of Static Problem}}
It will be shown that the static stiffness can be approximated by a finite sum of modal stiffnesses (see Figure \ref{fig:staticmodalstiffnessrelation} for a 1D case). Consider an elastic structure in static equilibrium under some applied load $\myvector{p}$. Supposing the structure is modelled by the finite element method, we end up with a system of linear equations for $3N$ degrees of freedom (DOFs) in 3D:
\begin{equation}
	\label{eq:SE}
	\mymatrix{K}\myvector{u} = \myvector{p}, \, \mymatrix{K} \in \mathbb{R}^{3N\times 3N},\, \myvector{u} \in \mathbb{R}^{3N}, \, \myvector{p} \in \mathbb{R}^{3N},
\end{equation}
where $\mymatrix{K}$ represents the stiffness matrix, $\myvector{u}$ is the vector of displacements, and $\myvector{p}$ is the load vector. Structure of $\myvector{u}$ and $\myvector{p}$ can be defined as follows
\begin{align}
	\myvector{u} = [u_{1,x},u_{1,y},u_{1,z},\ldots,u_{j,x},u_{j,y},u_{j,z},\ldots,u_{N,x},u_{N,y},u_{N,z}]^T,  \label{eq:displvector} \\
	\myvector{p} = [p_{1,x},p_{1,y},p_{1,z},\ldots,p_{j,x},p_{j,y},p_{j,z},\ldots,p_{N,x},p_{N,y},p_{N,z}]^T, \label{eq:loadvector}
\end{align}
where the first subscript of the entries of the vectors \eqref{eq:displvector} and \eqref{eq:loadvector} refers to the index of the point in the geometry and the second subscript determines direction of the displacement (or load); for example, $u_{j,y}$ relates to the displacement along axis $y$ at the point $j$.
\subsection{\revised{Formulation of Generalised Eigenvalue problem}}
The \revised{following} generalised eigenvalue problem \revised{is considered} in the form
\begin{equation}
	\mymatrix{K}\myvector{\phi}_k = \lambda_k \mymatrix{M} \myvector{\phi}_k,\, \myvector{\phi}_k \in \mathbb{R}^{3N},\,  \lambda_k \in \mathbb{R},  \, \mymatrix{M} \in \mathbb{R}^{3N\times 3N},\, k=1,\ldots,3N, \label{eq:GEP}
\end{equation}
where $\myvector{\phi}_k$ is the $k$'th eigenvector, $\lambda_k$ represents the \revised{corresponding} eigenvalue, and $\mymatrix{M}$ denotes the mass matrix.
Moreover, the eigenvalues $\lambda_k$ are numbered in \revised{non--decreasing} order, i.e., $\lambda_1 \le \lambda_2 \le \ldots \le \lambda_{3N}$. \revised{The $k$'th natural frequency of a structure $f_{k}$ and its relation with $\lambda_{k}$ is given by}
\begin{align}
	2\pi f_{k} = \sqrt{\lambda_{k}}.
\end{align}
The eigenvalues $\lambda_k$ define the diagonal \revised{matrices}
\begin{equation}
	\mymatrix{\Lambda}_{k} = \mathrm{diag}(\lambda_1,\lambda_2,\ldots,\lambda_{k}),\,  \mymatrix{\Lambda}_{k} \in \mathbb{R}^{k \times k},  \, k = 1,\ldots,3N. \label{eigmat}
\end{equation}
The eigenvectors $\myvector{\phi}_k$ define the \revised{matrices}
\begin{equation}
	\mymatrix{\Phi}_{k} = [\myvector{\phi}_1, \myvector{\phi}_2,\ldots,\myvector{\phi}_{k}],\quad k = 1,\ldots,3N \label{eigvecmat}
\end{equation}
the eigenvectors $\myvector{\phi}_k$ are scaled such that
\begin{equation}
	\mymatrix{\Phi}_{k}^T \mymatrix{M} \mymatrix{\Phi}_{k}  = \mymatrix{I}_{k},   \label{M-ortgohobality}
\end{equation}
where $\mymatrix{I}_{k}$ denotes the identity matrix. \revised{Therefore, it holds}
\begin{equation}
	\mymatrix{\Phi}_{k}^T \mymatrix{K} \mymatrix{\Phi}_{k} = \mymatrix{\Lambda}_{k}. \label{specdec}
\end{equation}
\revised{The mass matrix $\mymatrix{M}$ is considered in the form of its spectral decomposition} $\mymatrix{M} = \mymatrix{\Theta}\mymatrix{\Lambda}_M\mymatrix{\Theta}^T$, where \revised{ $\mymatrix{\Theta}$ denotes the matrix of the eigenvectors of $\mymatrix{M}$ such that} $\mymatrix{\Theta}^T\mymatrix{\Theta} = \mymatrix{I}_{3N}$ \revised{and $\mymatrix{\Lambda_M}$ represents the diagonal matrix of the corresponding eigenvalues. After some manipulation of Eq. \eqref{eq:SE}, one can obtain}
\begin{equation}
	\underbrace{\mymatrix{\Lambda}_M^{-1/2}\mymatrix{\Theta}^T\mymatrix{K}\mymatrix{\Theta}\mymatrix{\Lambda}_M^{-1/2}}_{\mymatrix{\Omega}^2} \mymatrix{\Lambda}_M^{1/2}\mymatrix{\Theta}^T \myvector{u} = \mymatrix{\Lambda}_M^{-1/2}\mymatrix{\Theta}^T\myvector{p}
	\label{eq:SEM}
\end{equation}
taking $\mymatrix{\Omega}^2 = \mymatrix{\Lambda}_M^{-1/2}\mymatrix{\Theta}^T\mymatrix{K}\mymatrix{\Theta}\mymatrix{\Lambda}_M^{-1/2}$, \revised{one can get}
\begin{equation}
	\mymatrix{\Omega}^2 \mymatrix{\Lambda}_M^{1/2}\mymatrix{\Theta}^T\myvector{u} = \mymatrix{\Lambda}_M^{-1/2}\mymatrix{\Theta}^T\myvector{p}.
	\label{eq:SEMsdx}
\end{equation}
\subsection{\revised{Truncated Spectral Decomposition}}
The truncated spectral decomposition of $\mymatrix{\Omega}^2$ \revised{is introduced}, such that the first $\ell$ (smallest) eigenvalues and the corresponding eigenvectors \revised{are} considered, i.e.,
\begin{equation}
	\mymatrix{\Omega}^2 \approx \mymatrix{\Omega}^2_{\ell} =  \mymatrix{\Psi}_{\ell}\mymatrix{\Lambda}_{\ell} \mymatrix{\Psi}_{\ell}^T, \quad  \mymatrix{\Psi}_{\ell}^T  \mymatrix{\Psi}_{\ell} = \mymatrix{I}_{\ell}, \quad \ell \revised{\lll} 3N.
	\label{eq:approxKM}
\end{equation}
The eigenvalues of $\mymatrix{\Omega}^2$ are the same as the eigenvalues of the generalised eigenvalue problem Eq.\eqref{eq:GEP}. The reduced model can be obtained by replacing $\mymatrix{\Omega}^2$ in Eq.\eqref{eq:SEMsdx} with $\mymatrix{\Omega}^2_{\ell}$ defined by Eq.\eqref{eq:approxKM} such that
\begin{equation}
	\mymatrix{\Psi}_{\ell} \mymatrix{\Lambda}_{\ell}  \underbrace{\mymatrix{\Psi}_{\ell}^T \mymatrix{\Lambda}_M^{1/2}\mymatrix{\Theta}^T}_{\Phi_{\ell}^{\dagger}} \myvector{u} \approx \mymatrix{\Lambda}_M^{-1/2}\mymatrix{\Theta}^T\myvector{p}.
	\label{eq:SEMsdapprox}
\end{equation}
Transformation of the original problem Eq.\eqref{eq:SE} to the Eq.\eqref{eq:SEMsdapprox} has been done in a similar fashion as \revised{transformation of} the generalised eigenvalue problem Eq.\eqref{eq:GEP} into the standard one. Therefore, it is \revised{straightforward} to give this definition
\begin{equation}
	\mymatrix{\Phi}_{\ell}^{\dagger} \equiv \mymatrix{\Psi}_{\ell}^T \mymatrix{\Lambda}_M^{1/2}\mymatrix{\Theta}^T,
	\label{pseudo}
\end{equation}
where the symbol $\mymatrix{\Phi}_{\ell}^{\dagger}$ denotes the \revised{Moore--Penrose  inverse} of $\mymatrix{\Phi}_{\ell}$, \revised{here expressed in the form of} its singular value decomposition (SVD). Let the $\ell$'th approximation of the vector of the generalised coordinates $\myvector{z}_{\ell}$ based on the vector $\myvector{u}$ be defined as
\begin{equation}
	\myvector{z}_{\ell} \equiv  \mymatrix{\Phi}_{\ell}^{\dagger} \myvector{u}, \label{gencoor_u}
\end{equation}
which can be seen as the linear least squares solution of the minimisation problem $\min\limits_{\myvector{z}_{\ell}} \| \mymatrix{\Lambda}_M^{1/2}\mymatrix{\Theta}^T\myvector{u} - \mymatrix{\Psi}_{\ell} \myvector{z}_{\ell} \|$. By using Eqs.\eqref{pseudo} and \eqref{gencoor_u} in Eq.\eqref{eq:SEMsdapprox} after some manipulation, the following can be obtained
\begin{equation}
	\myvector{z}_{\ell} \equiv
	\mymatrix{\Lambda}_{\ell}^{-1} \mymatrix{\Phi}_{\ell}^{\revised{\dagger}}  \myvector{p},
	\label{gencoor_p}
\end{equation}
which corresponds to the linear least squares solution of the minimisation problem $\min\limits_{\myvector{z}_{\ell}} \| \mymatrix{\Lambda}_M^{-1/2}\mymatrix{\Theta}^T\myvector{p}
	- \mymatrix{\Psi}_{\ell} \mymatrix{\Lambda}_{\ell} \myvector{z} \|$.
\subsection{\revised{Approximation of Static Compliance Tensor}}
A unit point load applied in three linearly independent directions in the Euclidean space is considered. Therefore, for the $j$'th point multiple right--hand--sides are \revised{considered} in the form
\begin{equation}
	\label{eq:kthload}
	\mymatrix{P}_j = \myvector{e}_j^{(N)} \otimes \mymatrix{I}_3, \, j=1,\ldots,N, \,  \mymatrix{P}_j \in \mathbb{R}^{3N \times 3}, \, \myvector{e}_j^{(N)} \in \mathbb{R}^{N}, \, \mymatrix{I}_3 \in \mathbb{R}^{3 \times 3},
\end{equation}
where $\myvector{e}^{(N)}_j$ is $j$'th Euclidean vector, symbol $\otimes$ represents the Kronecker
product. First, second and third columns of \eqref{eq:kthload} correspond to the unit load at $j$'th point in the directions $x$, $y$ and $z$, respectively. Solving Eq. \eqref{eq:SE} for the load configuration \eqref{eq:kthload} at $j$'th point leads to three solution vectors, which form the base $\mathbfrm{U}_{j} = [\mathbfrm{u}_{x}, \mathbfrm{u}_{y}, \mathbfrm{u}_{z}]$ that can then be used to compute the $3\times 3$ point compliance tensor at the $j$'th point
\begin{equation}
	\label{eq:pointstiffness}
	\mytensor{C}_{j}=\mymatrix{U}^{T}_{j}\mymatrix{K}\mymatrix{U}_{j}
\end{equation}
In virtue of Eqs. \eqref{eq:pointstiffness}, \eqref{gencoor_u} and \eqref{gencoor_p}, an approximation of the point static compliance can be obtained at $j$'th point as
\begin{equation}
	\label{eq:psc}
	\mytensor{C}_{j}\approxeq(\mymatrix{\Phi}_{l}\mymatrix{Z}_{j})^{T}\mymatrix{K}\mymatrix{\Phi}_{l}\mymatrix{Z}_{j}
\end{equation}
where the matrix of the generalised coordinates is $\mymatrix{Z}_{j} = [\myvector{z}_{j, x}, \myvector{z}_{j, y}, \myvector{z}_{j, z}]$.
If the modal space matrix $\mymatrix{\Phi}_{l}$ is reduced to a single $k$'th eigen-vector then the $k$'th modal point compliance tensor $\bar{\mytensor{C}}_{k, j}$ at $j$'th point will be obtained as
\begin{equation}
	\bar{\mytensor{C}}_{k, j}=(\myvector{\phi}_{k}\mymatrix{Z}_{j})^{T}\mymatrix{K}\myvector{\phi}_{k}\mymatrix{Z}_{j}
\end{equation}
Finally, the magnitude of either the static or modal point compliance in the direction of load $\myvector{t}$ at $j$'th point can be easily computed as
\begin{align}
	\norm{\mytensor{C}_{j}}=           & \myvector{t}^{T}_{j}\mytensor{C}_{j}\myvector{t}_{j}          \\
	\norm{\bar{\mytensor{C}}_{k, j}} = & \myvector{t}^{T}_{j}\bar{\mytensor{C}}_{k, j}\myvector{t}_{j} \\
	\norm{\mytensor{S}_{j}} =          & \frac{1}{\norm{\mytensor{C}_{j}}}                             \\
	\norm{\bar{\mytensor{S}}_{k, j}} = & \frac{1}{\norm{\bar{\mytensor{S}}_{k, j}}}
\end{align}
\subsection{The Smallest Stiffness Location and Direction}
\label{ss:smalleststiffnesslocationanddirection}
For practical reasons it is useful to reshape the vector $\myvector{\phi}_{k}$ as follows
\begin{align}
	\label{reshape}
	\myvector{\phi}^{rs}_{k} \equiv \left(\mymatrix{I}_{N} \otimes \sum\limits_{i=1}^3 (\myvector{e}_i^{(3)})^{\revised T}\right) \left( \left(\sum\limits_{i=1}^{N} \myvector{e}_i^{(N)} \otimes \mymatrix{I}_3\right) \circ \left(\myvector{\phi}_{k} \otimes \sum\limits_{i=1}^3 (\myvector{e}_i^{(3)})^{\revised{T}}\right) \right),\, \\
	\myvector{\phi}^{rs}_{k} \in \mathbb{R}^{N \times 3},\, k = 1,\ldots, 3N, \nonumber
\end{align}
where \revised{symbol} $\circ$ denotes the Hadamard (element--wise) product. The formula Eq.\eqref{reshape} corresponds to an operation $\myvector{\phi}^{rs}_{k} = \mathrm{reshape}(\myvector\phi_{k}, N, 3)$\footnote{The so called reshape operation on arrays is used in Numpy computational library \cite{oliphant2006guide}, which is used in this study.}.
Now, it is relatively easy to compute the vector of point--wise magnitudes of displacements for the $k$'th eigenvector and consequently pick the maximum displacement magnitude of this vector. The corresponding point index $j$ is  given by
\begin{equation}
	\label{eq:critical_point_j}
	j = \mathrm{arg\ max}\Big(\sqrt{\mathrm{diag}(\myvector{\phi}_{k}^{rs}(\myvector{\phi}^{rs}_{k})^{T})}\Big),
\end{equation}
where the $\mathrm{diag}(\cdot)$ operation applied on a matrix denotes its diagonal part. In this way analysing the maximum displacement magnitude resulting from the first $\ell$ smallest eigenvalues being considered is suggested. The goal is to find the smallest static/modal stiffness of a structure. Based on $k$'th eigenvector the corresponding point with the index $j$, its coordinates $\myvector{x}^{cp}_{j}$ and the direction $\myvector{t}_{j} = \myvector{\phi}_{k}^{rs}[j] / \norm{\myvector{\phi}_{k}^{rs}[j]}$ is obtained. Another way to motivate the solution for the critical point $j$, as defined by Eq.\eqref{eq:critical_point_j}, can be made as follows. Suppose an arbitrary point load $\myvector{p} = [0,0,0, \dots, P_x, P_y, P_z, \dots,\revised{0,0,0}]$. The norm (or the magnitude) of the modal compliance corresponding to the $k$'th eigenvector is defined as
\begin{equation}
	\norm{\bar{\mytensor{C}}_{k}} = z_k \myvector{\phi}_k^T\mymatrix{K}\myvector{\phi}_k z_k
\end{equation}
where $z_k = (\myvector{\phi}_k^T\myvector{p}) / \lambda_k$ and $\lambda_k = \myvector{\phi}_k^T\mymatrix{K}\myvector{\phi}_k$.
Then
\begin{gather}
	\norm{\bar{\mytensor{C}}_{k}} = \frac{\myvector{\phi}_k^T\myvector{p}}{\myvector{\phi}_k^T\mymatrix{K}\myvector{\phi_k}} (\myvector{\phi}_k^T\mymatrix{K}\myvector{\phi_k}) \frac{\myvector{\phi}_k^T\myvector{p}}{\myvector{\phi}_k^T\mymatrix{K}\myvector{\phi_k}} \\
	\norm{\bar{\mytensor{C}}_{k}} = \frac{(\myvector{\phi}_k^T\myvector{p})^2}{\myvector{\phi}_k^T\mymatrix{K}\myvector{\phi_k}} \\
	\norm{\bar{\mytensor{C}}_{k}} = \frac{(\myvector{\phi}_k^T\myvector{p})^2}{\lambda_k} \label{eq:findmax}
\end{gather}
It is clear from \eqref{eq:findmax} that for $\norm{\bar{\mytensor{C}}_{k}}$ to be maximised, one needs to pick the smallest magnitude eigenvalue $\lambda_k$ and choose the point load $\myvector{p}$ such that it maximises the inner product $\myvector{\phi}_k^T\myvector{p}$. Since $\myvector{p}$ must be a point load, one can find its critical application coordinate $\myvector{x}^{cp}$, or equivalently, the critical node index $j$, by  Eq.\eqref{eq:critical_point_j}. Nevertheless, the smallest eigen pair does not always yield the smallest static/modal stiffness due to the interaction between the point load and other eigenvectors. However, in the example cases of this study the smallest stiffness was always found to be within the first few eigen pairs.
\section{Semi--analytical Beam Stiffness Analysis}
\label{ss:beamstiffnessanalysis}
\begin{figure}[!bht]
	\centering
	\includegraphics[width=140mm]{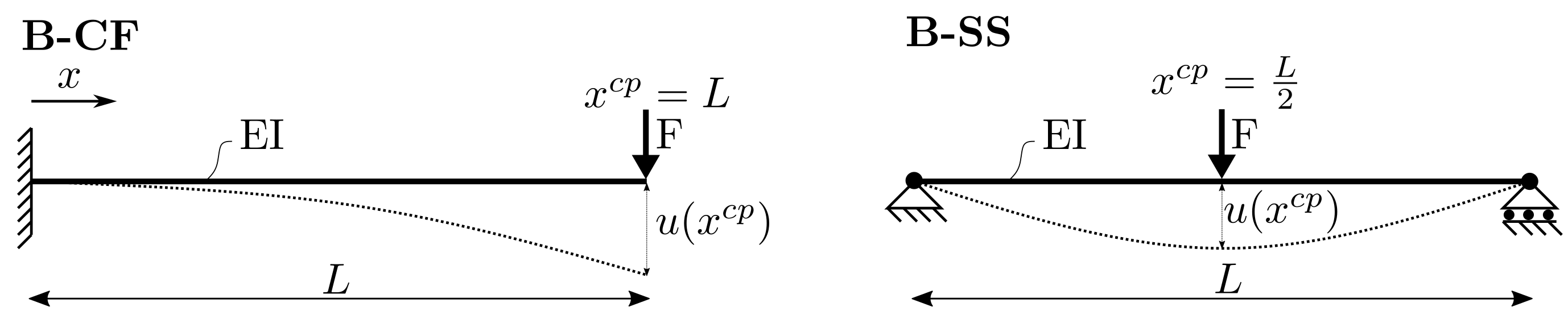}
	\caption{Beam boundary conditions and associated minimal stiffness deformation}
	\label{fig:beambcs}
\end{figure}

The relation between modal and static stiffness is demonstrated for the case of a simply supported beam (\textbf{B-SS})  \cite{wahyuni2010relationship}, it is further extended to the configuration of clamped--free boundary conditions (\textbf{B-CF}). Consider a simple beam whose mechanical response can be described by the classic Euler--Bernoulli beam theory:
\begin{equation}
	\label{eq:EB}
	\mathrm{E}\mathrm{I}\frac{\partial^{4} u}{\partial x^{4}} + \rho\mathrm{A}\frac{\partial^{2} u}{\partial t^{2}} = 0
\end{equation}
where $\mathrm{E}, \mathrm{I}, \rho, \mathrm{A}$ are Young's modulus, inertia moment, density and cross--section area respectively. The solution for the displacement $u$ can be found in the multiplicative form:
\begin{equation}
	u(x, t) = \Lambda(x)\Phi(t)
\end{equation}
where $\Lambda$ and $\Phi$ are temporal and time separable solutions of the Eq. \eqref{eq:EB}. Further separating the temporal and spatial parts of the solution and introducing a variable $\omega^{2}$, this must hold:
\begin{equation}
	\delta^{4} = \omega^{2} \frac{\rho\mathrm{A}}{\mathrm{E}\mathrm{I}}
\end{equation}
The spatial solution $\Lambda(x)$ can be expressed as an expansion:
\begin{equation}
	\Lambda(x) = \mathrm{C_{1}}sin(\delta x) + \mathrm{C_{2}}cos(\delta x) + \mathrm{C_{3}}sinh(\delta x) + \mathrm{C_{4}}cosh(\delta x)
\end{equation}
The constants $\mathrm{C_{1}},..,\mathrm{C_{4}}$ can be found with suitable boundary conditions defined. Here the description of the solution for configurations \textbf{B-SS} and \textbf{B-CF} is ommited and instead it is suggested the reader the appropriate literature with precise description on how to compute coefficients $\mathrm{C_{1}},..,\mathrm{C_{4}}$ (for example in \cite{avcarM2014}). The characteristic equations for $\delta L$ for both configurations \textbf{B-CF}, \textbf{B-SS} are:
\begin{align}
	\label{eq:waveequations}
	cos(\delta L)cosh(\delta L) = -1, \\
	sin(\delta L) = 0,
\end{align}
which are non--linear wave functions, for which the $k$'th solution is known to be:
\begin{align}
	\label{eq:ss}
	\delta_{k}^{B-CF} L\approx & \frac{(2k - 1 + e)\pi}{2}, \\
	\delta_{k}^{B-SS} L\approx & k\pi,
\end{align}
where $k$, is the $k$'th mode shape and $e$ is an expansion strongly approaching zero: $0.386, 0.01,..$. Now consider static bending stiffness for both configurations:
\begin{align}
	\label{eq:staticstiffnesses}
	\mathbb{S}^{B-CF} = \frac{3\mathrm{E}\mathrm{I}}{L^{3}}, \\
	\mathbb{S}^{B-SS} = \frac{48\mathrm{E}\mathrm{I}}{L^{3}},
\end{align}
apparently, the maximum for \textbf{B-CF} is at the right end $x^{cp}=L$ and the middle for \textbf{B-SS} $x^{cp}=L/2$ (see Figure \ref{fig:beambcs}).
Further consider $k$'th solution for \textbf{B-CF}, which is described as trigonometric expansion in a form:
\begin{align}
	\label{eq:BCFsolution}
	\sigma_{k} =  & \frac{\cos(\delta_{k} L) + \cosh(\delta_{k} L)}{\sin(\delta_{k} L) + \sinh(\delta_{k} L)},       \\
	\phi_{k}(x) = & \cosh(\delta_{k} x) - \cos(\delta_{k} x) + \sigma_{k}(\sin(\delta_{k} x) - \sinh(\delta_{k} x)),
\end{align}
where $\delta_{k}$ represents solution of associated $k$'th wave equation \eqref{eq:waveequations}. The $k$'th eigenvalue can be computed from relation $\lambda_{k} = \omega_{k}^2 M_{k}$ where the $k$'th modal mass $M_{k}$ is computed as:
\begin{align}
	M_{k} =      & \rho\mathrm{A}\int_{0}^{L}\phi_{k}^{2}(x)\intdiff{x} \label{eq:BCFintegration}, \\
	\omega_{k} = & \delta_{k}^{2}\sqrt{\frac{\mathrm{EI}}{\rho A}} \label{eq:freqexpression}.
\end{align}
Finally, the $k$'th eigenvalue is given as:
\begin{equation}
	\lambda_{k} = \kappa_{k}(\delta_{k})\mathrm{E}\mathrm{I}L,
\end{equation}
where the constant $\kappa_{k}$ represents irrational number from integration \eqref{eq:BCFintegration} and the $k$'th solution \eqref{eq:BCFsolution} for $\delta_{k}$. The static stiffness can be approximated by a modal stiffness in a form, which is a 1D simplification of \eqref{eq:psc}:
\begin{equation}
	\frac{1}{\mathbb{S}}\approxeq\sum_{k=1}^{l}\phi_{k}(x^{cp})z_{k} = \sum_{k=1}^{l}\frac{\phi_{k}^{2}(x^{cp})}{\lambda_{k}}.
\end{equation}
Comparing the resultant modal stiffness with the analytical and removing common terms, the following relation is obtained:
\begin{equation}
	\frac{\mathbb{S}}{\mathbb{S}^{B-CF}}=\frac{\frac{\phi_{1}^{2}(x^{cp})}{\kappa_{1}}+\frac{\phi_{2}^{2}(x^{cp})}{\kappa_{2}}+...+\frac{\phi_{n}^{2}(x^{cp})}{\kappa_{n}}}{n} \geq 1, n=l.
\end{equation}
For a sufficient number of modes, the ratio above tends to reach unity from above since the constants $\kappa$ and the squares of the shape vectors are always positive. The limit of the series also implies that for a finite number of modes included the approximate stiffness $\mytensor{S}$ is always higher than the static stiffness at the critical point $x^{cp}$. Another important fact is that the modal stiffness defined by only the first modal shape can approximate the static stiffness with sufficient accuracy (for example, $\frac{\bar{\mytensor{S}}_{1}}{\mytensor{S}^{B-CF}}=1.029$). Nevertheless, this is only possible for a static load and boundary conditions, which produce a ``\emph{similar}'' deformation shape as the corresponding shape vector. This approach can be applied for the beam with configuration \textbf{B-SS}. The resultant expression is
\begin{equation}
	\frac{\mathbb{S}}{\mathbb{S}^{B-SS}} = \frac{\pi^{4}}{96(1 + 0 + \frac{1}{3^{4}} + ...+ 0 + \frac{1}{n^{4}})} \geq 1, n=l.
\end{equation}
The constant $\kappa$ is now given precisely as $n^{4} > 0$ for even $n$ (unsymmetrical modes). Again, if only the first modal shape is employed, the ratio $\frac{\bar{\mytensor{S}}_{1}}{\mytensor{S}^{B-SS}}=1.0147$ is obtained.
\section{Finding the Critical Loading Point by \revised{Numerical Optimisation}}
\label{s:gradientbasedmethodtofindcriticalloadingpoint}
\subsection{Maximum Compliance Deformation}
The objective is to find the minimum stiffness of an elastic solid under some prescribed Dirichlet boundary conditions. The minimum stiffness deformation is one where a unit-magnitude external force does maximal work; or, equivalently, it is the force that causes maximal strain energy to be stored in the elastic solid.
\subsection{Problem Statement}
The aim is to find the point force that maximises the strain energy stored in an elastic solid. However, dealing with point forces in a numerical optimisation context is not practical; thus, we describe the force more generally as a body force field defined on the whole domain. The hope is that the optimal solution will result in the localisation of this body force into what could be considered as a point load. The objective is to find a unit--magnitude body-force $\mathbfrm{f}^\star \in \{L_2(\Omega) | \int_\Omega ||\mathbfrm{f}||_2\intdiff{x} = 1\}$ that maximises the elastic strain energy $U(\mathbfrm{u}(\mathbfrm{f}))$ of the deformation $\mathbfrm{u}(\mathbfrm{f})$. The optimisation problem can be summarized as follows
\begin{eqnarray}
	& \mathbfrm{f}^\star = \argmax\limits_{\mathbfrm{f} \in L_2(\Omega)} U (\mathbfrm{u}(\mathbfrm{f})) \\
	\mbox{subject to} \quad
	& F(\mathbfrm{u}, \mathbfrm{f}; \mathbfrm{v}) = 0 \quad \forall\mathbfrm{v}, \label{eq:EquilibriumVariationalForm}\\
	& \int_\Omega ||\mathbfrm{f}||_2\intdiff{x} = 1 \label{eq:ConstraintUnitForce} \quad \mathbfrm{f} \in L_2(\Omega)
	\label{eq:ForceConstraintL1Norm}
\end{eqnarray}
Equation \eqref{eq:EquilibriumVariationalForm} is the variational form of static equilibrium that can be generally written as
\begin{equation} \label{eq:VariationalProblemOfEquilibrium}
	F(\mathbfrm{u}, \mathbfrm{f}; \mathbfrm{v}) := \partial_u U(\mathbfrm{u})[\mathbfrm{v}] - \partial_u W(\mathbfrm{u}, \mathbfrm{f})[\mathbfrm{v}] = 0 \quad  \forall\mathbfrm{v},
\end{equation}
where $W(\mathbfrm{u}, \mathbfrm{f})$ is the external work done by the force $\mathbfrm{f}$ and $\mathbfrm{v}$ is any admissible virtual displacement field variation. The notation $\partial_a (\cdot) [b]$ denotes a directional derivative with respect to an argument $a$ in the direction of $b$. Equation \eqref{eq:VariationalProblemOfEquilibrium} implicitly defines the function--like relationship between $\mathbfrm{f}$ and $\mathbfrm{u}$, i.e. $\mathbfrm{f}\rightarrow \mathbfrm{u}(\mathbfrm{f})$, since $\mathbfrm{u}(\mathbfrm{f})$ is required to satisfy \eqref{eq:VariationalProblemOfEquilibrium} for any admissible $\mathbfrm{f}$. It is important to highlight that the force constraint \eqref{eq:ForceConstraintL1Norm} is an $L_1$--norm constraint; it specifies that the total amount of force applied in the solid must be one unit.
\subsection{Solution Method}
\label{ss:solutionmethod}
The force $\mathbfrm{f}$ that maximises $U(\mathbfrm{u}(\mathbfrm{f}))$ can be solved for by using a gradient-based solution strategy. To this end, we require the Fr\'echet derivative of $U(\mathbfrm{u}(\mathbfrm{f}))$ with respect to $\mathbfrm{f}$
\begin{equation} \label{eq:CostGradientUsingDirectMethod}
	D_f U = \partial_u U \, D_f \mathbfrm{u},
\end{equation}
where $\partial_u U$ is the Fr\'echet derivative with respect to $\mathbfrm{u}$. $D_f \mathbfrm{u}$ can be obtained by differentiating the variational form \eqref{eq:VariationalProblemOfEquilibrium} of the static equilibrium equations
\begin{gather}
	\label{eq:DirectionalDerivativeOfDisplacementField}
	\partial_u F(\mathbfrm{u}, \mathbfrm{f}, \mathbfrm{v})[\delta \mathbfrm{u}] + \partial_f F(\mathbfrm{u}, \mathbfrm{f}, \mathbfrm{v})[\delta \mathbfrm{f}] = 0 \quad \forall \mathbfrm{v} \\
	D_f \mathbfrm{u} = -\partial_u F ^{-1}(\mathbfrm{u}, \mathbfrm{f}, \mathbfrm{v})[\mathbfrm{v}'] \, \partial_f F(\mathbfrm{u}, \mathbfrm{f}, \mathbfrm{v}) \quad \forall \mathbfrm{v}, \mathbfrm{v}'
	\label{eq:FrechetDerivativeOfDisplacementField}
\end{gather}
where $\partial_u F ^{-1}(\mathbfrm{u}, \mathbfrm{f}, \mathbfrm{v})[\mathbfrm{v}']$ (subsequently written as $\partial_u F^{-1}$) denotes an inverse--like operator of the bilinear form $\partial_u F(\mathbfrm{u}, \mathbfrm{f}, \mathbfrm{v})[\mathbfrm{v}']$ (subsequently written as $\partial_u F$) in which $\mathbfrm{v}$ and $\mathbfrm{v}'$ are the first and second arguments associated with the admissible virtual displacement field variations. Substituting \eqref{eq:FrechetDerivativeOfDisplacementField} into \eqref{eq:CostGradientUsingDirectMethod} gives
\begin{equation} \label{eq:CostGradientUsingDirectMethodExpanded}
	D_f U = -\partial_u U \, \partial_u F ^{-1} \, \partial_f F
\end{equation}
Note that the inverse $\partial_u F ^{-1}$ need not be computed as only the effect of $\partial_u F ^{-1} \partial_f F$ or, alternatively, $\partial_u U \, \partial_u F ^{-1}$ is required. This presents a choice whether to compute the former or the latter term. As the dimension of $\partial_f F$ (matrix-like) is usually much larger than that of $\partial_u U$ (vector--like), it is computationally advantageous to compute $\partial_u U \, \partial_u F ^{-1}$. This leads to the so-called adjoint method of computing $D_f U$:
\begin{gather} \label{eq:CostGradientUsingAdjointMethod}
	D_f U = \partial_f F(\mathbfrm{u}, \mathbfrm{f}, \mathbfrm{\lambda}) \\
	\mathbfrm{\lambda} = - \mathrm{adj} \left(\partial_u F\right)^{-1} \partial_u U
	\label{eq:AdjointEquation}
\end{gather}
where $\mathrm{adj} \left(\partial_u F\right):= \partial_u F(\mathbfrm{u},\mathbfrm{f}; \mathbfrm{v}')[\mathbfrm{v}]$ is the adjoint of $\partial_u F(\mathbfrm{u}, \mathbfrm{f}; \mathbfrm{v})[\mathbfrm{v}']$ and $\mathbfrm{\lambda}$ is the so--called adjoint variable. In the context of linear elasticity, $\partial_u F(\mathbfrm{u}, \mathbfrm{f}; \mathbfrm{v})[\mathbfrm{v}']$ is always symmetric with respect to the arguments $\mathbfrm{v}$ and $\mathbfrm{v}'$; hence, the adjoint of $\partial_u F$ will be itself. Consequently, \eqref{eq:AdjointEquation} simplifies to $\mathbfrm{\lambda} = -\partial_u F^{-1} \partial_u U$.
The $L_1$--norm constraint \eqref{eq:ForceConstraintL1Norm} poses a numerical difficulty in applying the gradient-based solution strategy in the maximisation of $U(\mathbfrm{u}(\mathbfrm{f}))$ because the gradient of the constraint becomes ill--defined as the body force approaches zero magnitude point--wise; specifically,
\begin{equation}
	\partial_{f_i} \int_\Omega ||\mathbfrm{f}||_2\intdiff{x} = \int_\Omega \frac{f_i}{||\mathbfrm{f}||_2}\intdiff{x}
\end{equation}
Intuitively, the incremental solution for $\delta\mathbfrm{f}$ that maximises $\delta_f U = D_f U \, \delta\mathbfrm{f}$ under the $L_1$-norm constraint will be one that localises at a single point where the absolute value of $D_f U (\mathbfrm{x})$ for $\mathbfrm{x} \in\Omega$ is largest. However, such $\delta\mathbfrm{f}$ is a spatially non--smooth change in $\mathbfrm{f}$. The negative effect of such $\delta\mathbfrm{f}$ is that it tends to cause $\mathbfrm{f}$ to localise prematurely at that same point with subsequent iterations. Such a solution may not be globally optimal. For the numerical solution algorithm to be successful, a solution increment $\delta\mathbfrm{f}$ must be regularised so that the evolutions of $\mathbfrm{f}$ and $U(\mathbfrm{f})$ may be spatially smooth during the iterative solution process. In proposed implementation, the gradual and spatially smooth evolution of $\mathbfrm{f}$ was achieved by gradually thresholding $\mathbfrm{f}$. Specifically, the solution algorithm is as follows
\begin{enumerate}
	\label{items:algorithm}
	\item For the body-force threshold $T^i\in (0, 1)$
	      \begin{enumerate}
		      \item Solve the equilibrium problem \eqref{eq:VariationalProblemOfEquilibrium} based on the current  $\mathbfrm{f}^k$
		      \item Compute the energy gradient $D_f U^k$ using \eqref{eq:CostGradientUsingAdjointMethod} and \eqref{eq:AdjointEquation}
		      \item Compute the solution advance direction $\hat{\mathbfrm{f}}^k = D_f U^k / ||D_f U^k||$
		      \item Compute a tentative new solution $\tilde{\mathbfrm{f}}^{k+1} = \mathbfrm{f}^k + \hat{\mathbfrm{f}}^k \Delta$
		      \item Compute the body-force threshold level $\tilde f_\mathrm{th-min}^{k+1} = \tilde f_\mathrm{max}^{k+1} T^i $ where \\ ${\tilde f^{k+1}_\mathrm{max} = \max \limits_{x \in \Omega} ||\tilde{\mathbfrm{f}}^{k+1}(\mathbfrm{x})||_2}$
		      \item Compute the thresholded solution \\ ${\tilde{\mathbfrm{f}}_\mathrm{th}^{k+1}(\mathbfrm{x}) = \{\tilde{\mathbfrm{f}}^{k+1}(\mathbfrm{x}) \; \mbox{if} \; ||\tilde{\mathbfrm{f}}^{k+1}(\mathbfrm{x})||_2 > \tilde f_\mathrm{th-min}^{k+1} \; \mathrm{for} \; \mathbfrm{x} \in \Omega, \; \mbox{otherwise} \; 0\}}$
		      \item Enforce constraint \eqref{eq:ForceConstraintL1Norm} on $\tilde{\mathbfrm{f}}_\mathrm{th}^{k+1}$ to obtained the solution for $k+1$ \\
		            ${\mathbfrm{f}^{k+1} = \tilde{\mathbfrm{f}}_\mathrm{th}^{k+1} / \int_{\Omega} ||\tilde{\mathbfrm{f}}_\mathrm{th}^{k+1}||_2\intdiff{x}}$
	      \end{enumerate}
	\item If $T^i$ is below a user--prescribed maximum value, increment the body--force threshold  $T^{i+1} = T^i + \Delta T$ and continue from {\tt(a)}; otherwise, stop.
\end{enumerate}
The algorithm is implemented in computational library FEniCS \cite{logg2012automated} and it is available on author's \href{https://github.com/danassutula/critical_loading_case}{GitHub repository}.
\section{Computational Models}
\label{s:computationalmodels}
\begin{figure}[!bht]
	\centering
	\includegraphics[width=140mm]{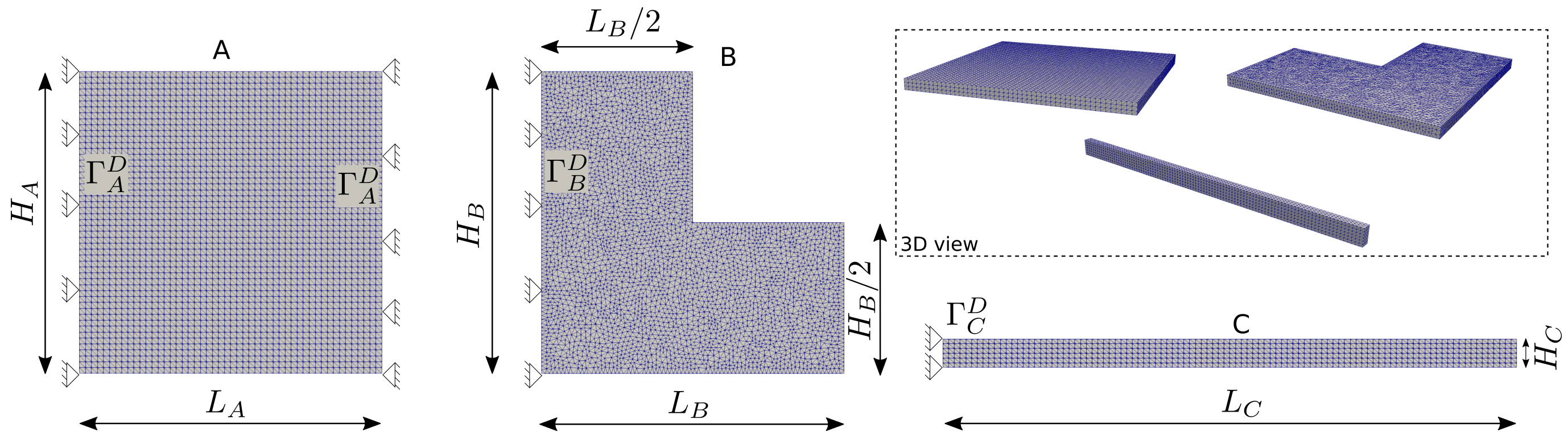}
	\caption{Three computational meshed domains with fixed edges defined by Dirichlet boundary conditions $\Gamma_{D}$. \revised{The dimensions of models are $H_A=L_A=10$ mm, $H_B=9, L_B=10$ mm and $L_C=20, H_C=1$ mm}. Mesh size is fixed for all models and is $L_{A}/50,L_{B}/50,L_{C}/50$.}
	\label{fig:benchmarkmodels}
\end{figure}

\begin{figure}[!bht]
	\centering
	\includegraphics[width=140mm]{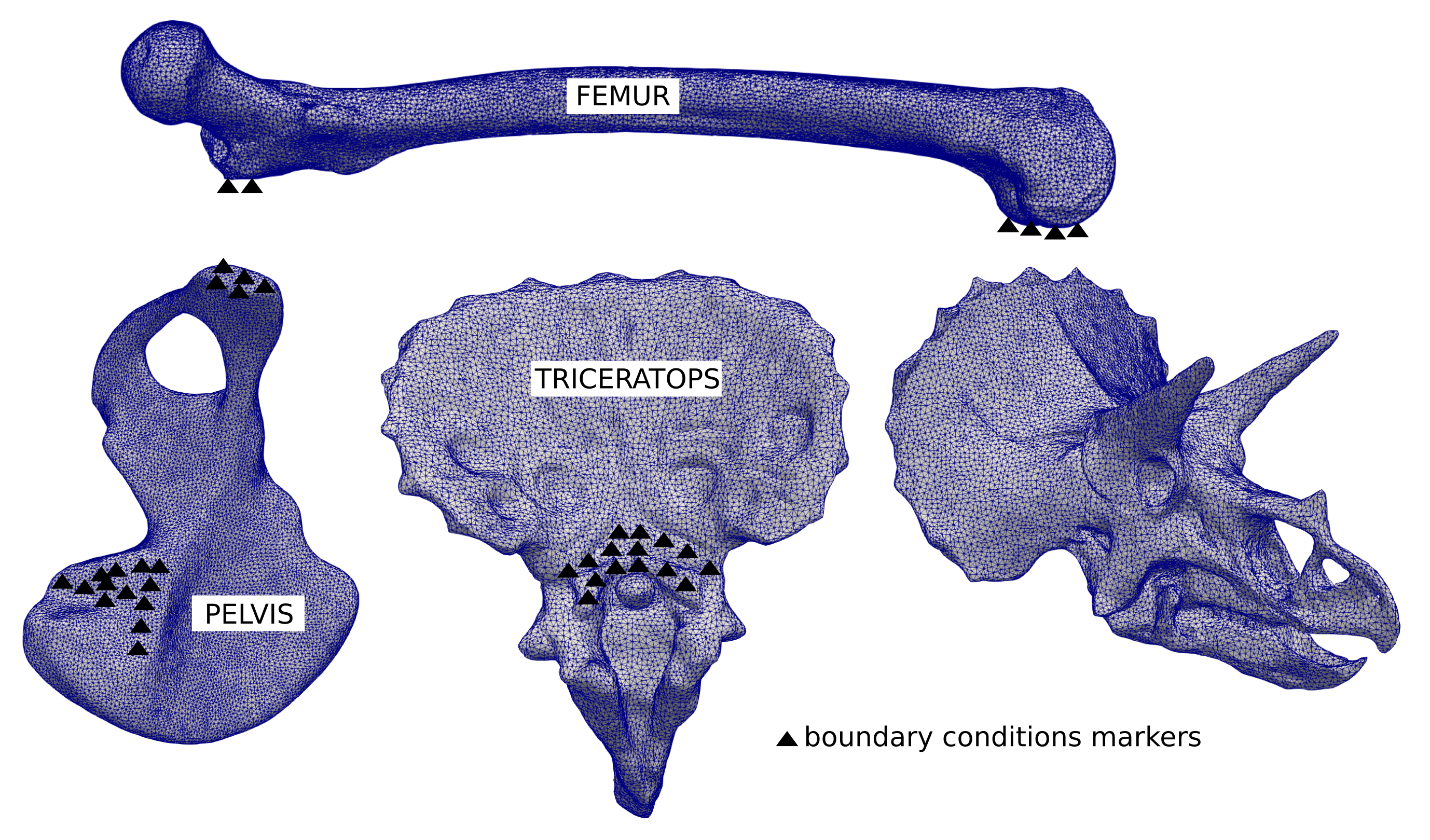}
	\caption{Complex biological shapes. The optimal mesh element characteristic size is 1 mm for all models (the TetWild mesher \cite{hu2018tetrahedral}).}
	\label{fig:benchmarkmodelsfunny}
\end{figure}

The static and modal stiffnesses are assessed in the following benchmarks. The smallest stiffness is computed by spectral approach as well as by the gradient--optimisation method introduced in Section \ref{s:gradientbasedmethodtofindcriticalloadingpoint}. Three simple examples, which are shown in Figure \ref{fig:benchmarkmodels}, and several more complex biological shapes of bones as shown in Figure \ref{fig:benchmarkmodelsfunny} are introduced. All models are 0.5 mm thick. The material is defined by Young's modulus E=10 GPa for structures A, B and C and E=5 GPa for biological shapes. The Poisson's ratio is $\nu=0.3$ for all examples.
The computational models assume linear elasticity. Under the small--strain assumption, the infinitesimal strain tensor $\mathbfrm{\varepsilon}$ is defined as:
\begin{equation}
	\mathbfrm{\varepsilon} = \frac{1}{2}(\nabla\mathbfrm{u} + \nabla^{T}\mathbfrm{u}).
\end{equation}
The constitutive relation for the stress tensor is expressed as:
\begin{equation}
	\mathbfrm{\sigma} = \mathbfrm{C}:\mathbfrm{\varepsilon},
\end{equation}
where the elasticity tensor $\mytensor{C}$ for an isotropic material can be written as:
\begin{equation}
	\label{eq:ETT}
	C_{ijkl} = \frac{E\mu}{(1 + \mu)(1 - 2\mu)}\delta_{ik}\delta_{kl} + \frac{E}{1+\mu}\delta_{il}\delta_{jk}
\end{equation}
The static equilibrium solution of the linear elasticity problem in domain $\Omega$ is given by the solution to the weak form (which is a simplification of \eqref{eq:VariationalProblemOfEquilibrium}):
\begin{align}
	\label{eq:NF}
	\int_{\Omega}\mathbfrm{\varepsilon(u)}:\mathbfrm{\sigma(v)}\intdiff{\Omega} = 0, \forall\mathbfrm{v}
\end{align}
Moreover, the stiffness and mass matrices needed for solving the generalised eigen value problem are defined as:
\begin{align}
	\label{eq:matrices}
	\mathbfrm{K} = & \int_{\Omega}\mathbfrm{\varepsilon(\mathbfrm{u})}:\mathbfrm{\sigma(\mathbfrm{v})}\intdiff{\Omega} \\
	\mathbfrm{M} = & \rho\int_{\Omega}\mathbfrm{u}\cdot\mathbfrm{v}\intdiff{\Omega}
\end{align}
The homogeneous Dirichlet boundary conditions are applied at the algebraic level, i.e. by modifying the stiffness matrix (refer to the FEniCS library manual \cite{logg2012automated}). Volume force $\myvector{f}$, displacement $\myvector{u}$ and their variations were discretised by the usual piecewise linear finite elements.
\section{Interpretation of Free--free Boundary Conditions}
\begin{figure}[!bht]
	\centering
	\includegraphics[width=140mm]{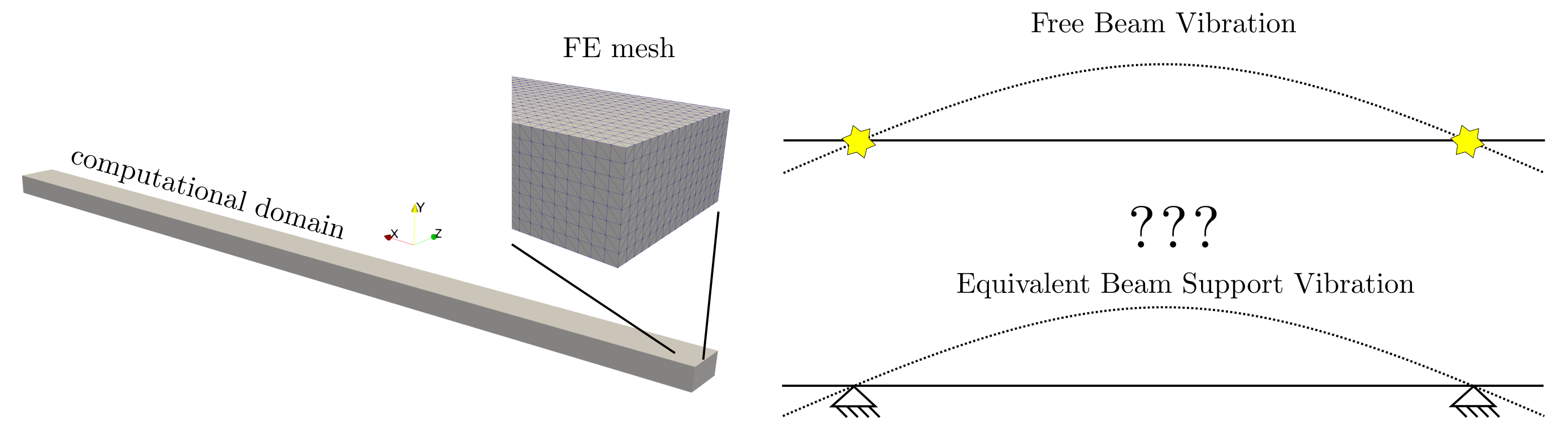}
	\caption{Computational model of a beam with free and equivalent boundary conditions.}
	\label{fig:femodelbcconfigurations}
\end{figure}

So far, it has been shown that static and modal stiffness is in close relation for a given set of boundary conditions. Nevertheless, there is a solution with free--free boundary conditions, for which the interpretation of the modal stiffness is not so obvious. The previously discussed relation of static and modal stiffness was demonstrated via simple analytical solutions to the transverse vibration problem of a slender beam. For a deeper exploration of the static--modal relation, a more complex computational model of a beam is presented. A second--order solid finite element approximation of the geometry and displacement field was used. The geometry and boundary conditions are shown in Figure \ref{fig:femodelbcconfigurations}.
The five eigenvalues were computed and equivalent boundary conditions where imposed based on the eigenvectors $\myvector{\phi}$. At the zero--displacement locations (called nodes) the stiffness is considered infinite\revised{, hence those nodes} present the singular points of the modal stiffness. The locations with approximate zero modal displacements are identified and taken as points for applying kinematic boundary conditions.
\section{Results}
\label{s:results}
\subsection{Spectral Analysis}
\label{ss:spectralanalysis}
\begin{figure}[!bht]
	\centering
	\includegraphics[width=1\linewidth]{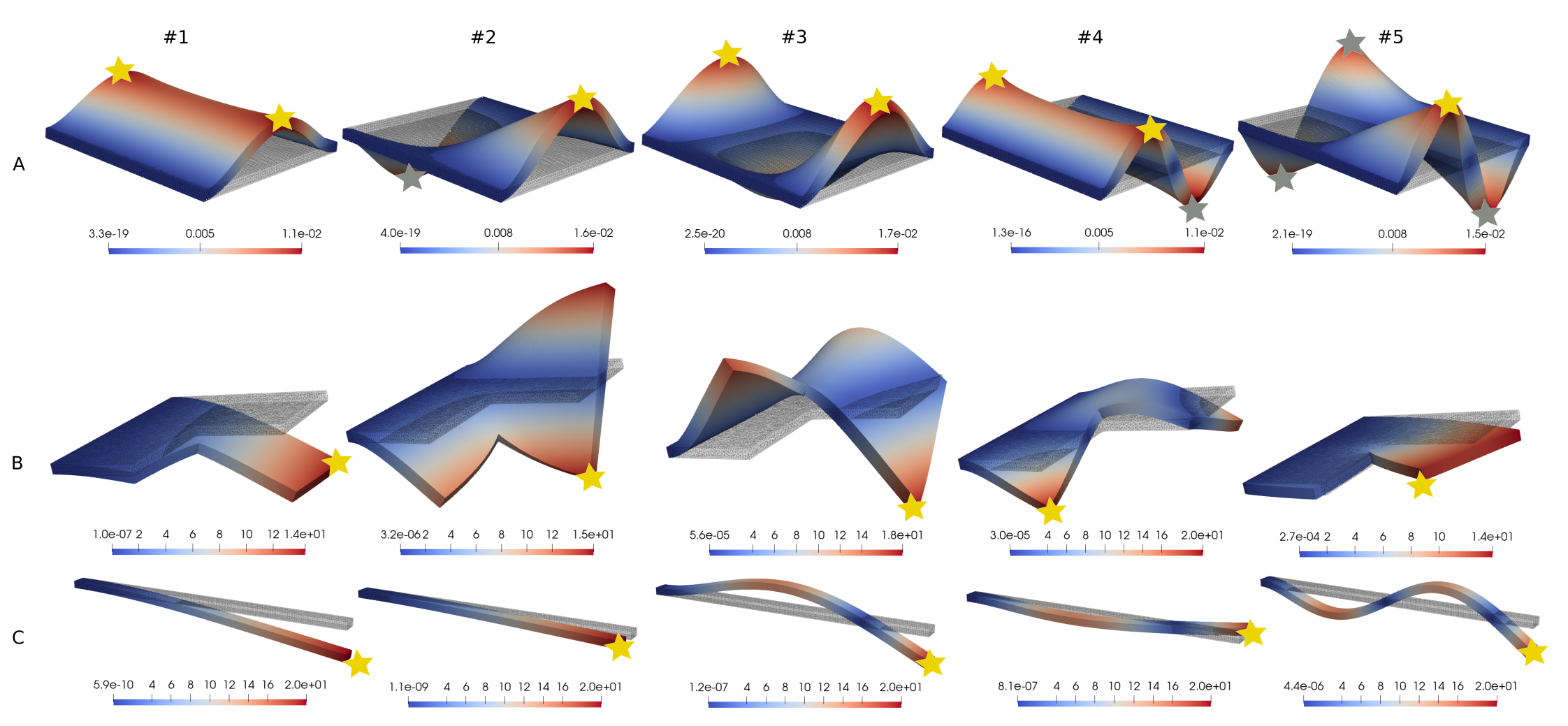} 
	\caption{\revised{Magnitudes [mm] of modal} shapes of three computational models. Yellow stars represent the maximum value of displacement magnitude at a point $\mathbfrm{x}^{cp}$. The grey stars represent symmetric/antisymmetric counterparts of the maximum values.}
	\label{fig:shapeextremes}
\end{figure}

\begin{figure}[!bht]
	\centering
	\includegraphics[width=1\linewidth]{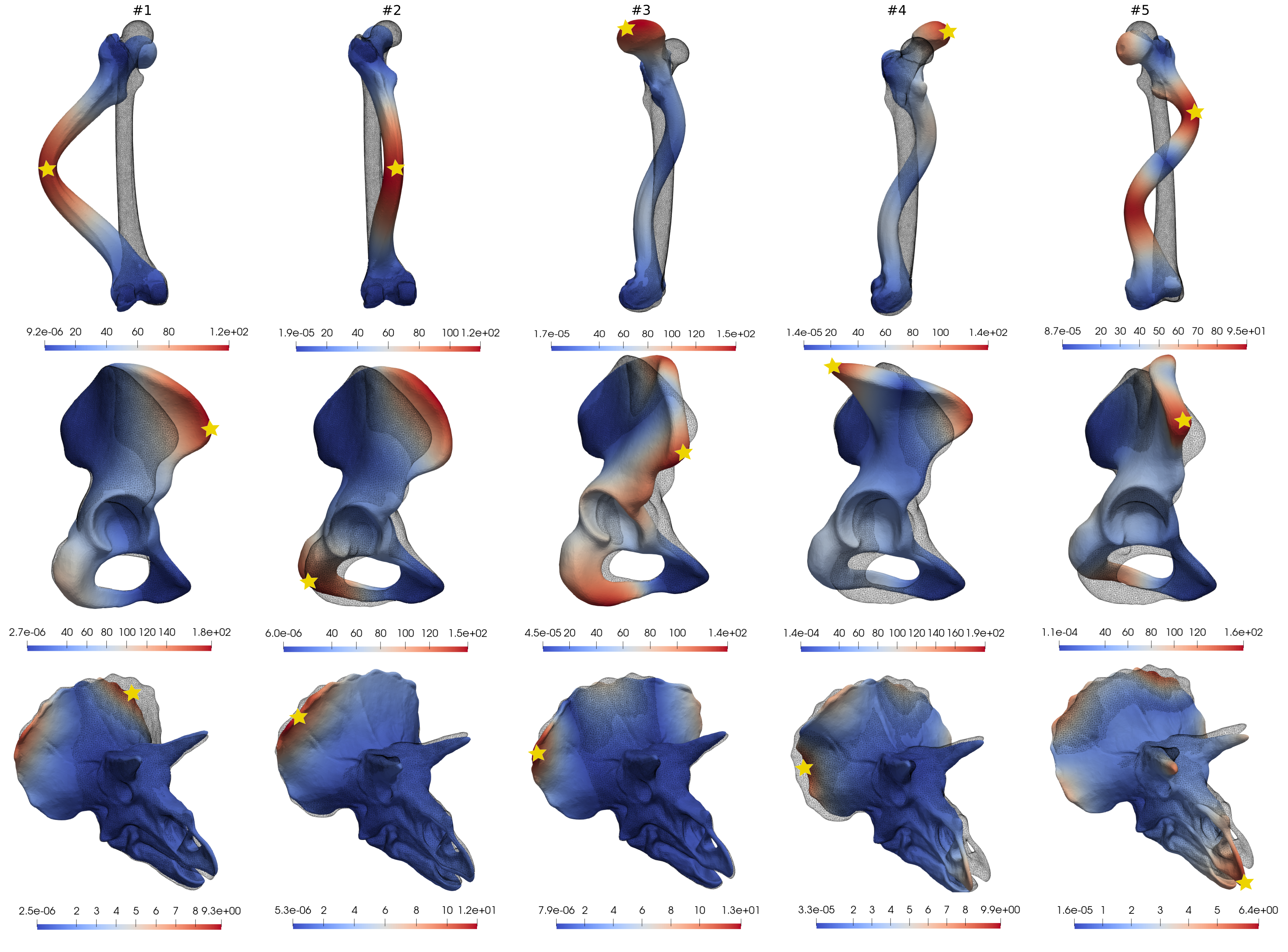} 
	\caption{\revised{Magnitudes [mm] of} modal shapes of three of complex biological shapes. Yellow stars represent the maximum value of displacement magnitude at a point $\mathbfrm{x}^{cp}$.}
	\label{fig:complegeomshapes}
\end{figure}

The five eigenvectors of given tested structures are shown in Figures \ref{fig:shapeextremes} and \ref{fig:complegeomshapes}. The shapes reflect the symmetry/asymmetry of geometry as well as shape indeed. The shapes of the squared plate (A) live in out of the plane direction and are closely related to bending stiffness. The other models (B, C) also exhibit the bending characteristic in the shapes. The biological shapes have complex deformation shapes except for the model FEMUR, which is quite similar to a beam.
The modal stiffness, static stiffness and natural frequencies for each structure are given in Tables \ref{tab:eigenstiffness} and \ref{tab:eigenstiffnesscomplexshapes}. Due to the rescaling by the amplitude squared factor, the modal stiffness does not necessarily follow the same order as the natural frequencies do. The static stiffness \revised{does not follow the order of natural frequencies} as well. Moreover, it depends on the point location and direction, which can be the same for multiple modes, see for example structure C and its static stiffness for odd and even eigenvectors, which have a maximum in the same location and direction. Similar patterns of modal and static stiffness behaviours are seen for biological shapes \revised{tested}.
\begin{table}[!bht]
	\caption{\revised{The static and dynamic properties} of analysed models A,B,C.}
	\label{tab:eigenstiffness}
	\centering
	\begin{adjustbox}{max width=\textwidth}
		\begin{tabular}{|c|ccc|ccc|ccc|}
			\hline
			        & \multicolumn{3}{c|}{Modal stiffness [N/mm]} & \multicolumn{3}{c|}{Static stiffness [N/mm]} & \multicolumn{3}{c|}{Natural frequency [Hz]}                                                     \\
			Mode \# & A                                           & B                                            & C                                           & A      & B       & C    & A      & B      & C     \\
			\hline
			1       & 1699.33                                     & 28.21                                        & 0.41                                        & 505.49 & 24.71   & 0.39 & 52.59  & 12.21  & 2.01  \\
			2       & 1033.36                                     & 264.11                                       & 1.61                                        & 505.11 & 30.52   & 1.56 & 61.92  & 41.27  & 4.11  \\
			3       & 2347.51                                     & 425.18                                       & 15.82                                       & 507.21 & 30.33   & 0.39 & 101.11 & 56.66  & 12.64 \\
			4       & 12040.52                                    & 528.84                                       & 61.91                                       & 885.28 & 162.41  & 1.58 & 141.87 & 73.13  & 25.02 \\
			5       & 6990.61                                     & 3123.95                                      & 122.64                                      & 885.33 & 2951.17 & 0.39 & 154.71 & 123.49 & 35.21 \\
			\hline
		\end{tabular}
	\end{adjustbox}
\end{table}
\begin{table}[!bht]
	\caption{\revised{The static and dynamic properties} of analysed complex biological shape models}
	\label{tab:eigenstiffnesscomplexshapes}
	\centering
	\begin{adjustbox}{max width=\textwidth}
		\begin{tabular}{|c|ccc|ccc|ccc|}
			\hline
			        & \multicolumn{3}{c|}{Modal stiffness [N/mm]} & \multicolumn{3}{c|}{Static stiffness [N/mm]} & \multicolumn{3}{c|}{Natural frequency [Hz]}                                                                    \\
			Mode \# & FEMUR                                       & PELVIS                                       & TRICERATOPS                                 & FEMUR   & PELVIS & TRICERATOPS & FEMUR   & PELVIS  & TRICERATOPS \\
			\hline
			1       & 919.98                                      & 403.32                                       & 3908.97                                     & 800.15  & 283.53 & 2554.59     & 571.80  & 591.16  & 92.47       \\
			2       & 1456.78                                     & 1076.46                                      & 2621.87                                     & 1253.47 & 354.01 & 1191.37     & 726.24  & 790.659 & 97.29       \\
			3       & 1369.64                                     & 4032.89                                      & 3971.91                                     & 962.91  & 524.79 & 1114.16     & 901.66  & 1389.95 & 129.27      \\
			4       & 2500.02                                     & 3230.99                                      & 9432.41                                     & 1248.27 & 816.75 & 2394.18     & 1075.29 & 1729.45 & 152.73      \\
			5       & 6592.80                                     & 5043.08                                      & 27589.96                                    & 1385.53 & 903.52 & 3073.32     & 1230.14 & 1809.14 & 170.44      \\
			\hline
		\end{tabular}
	\end{adjustbox}
\end{table}
\subsection{Optimisation}
\label{ss:optimizationresults}
\begin{figure}[!bht]
	\centering
	\includegraphics[width=1\linewidth]{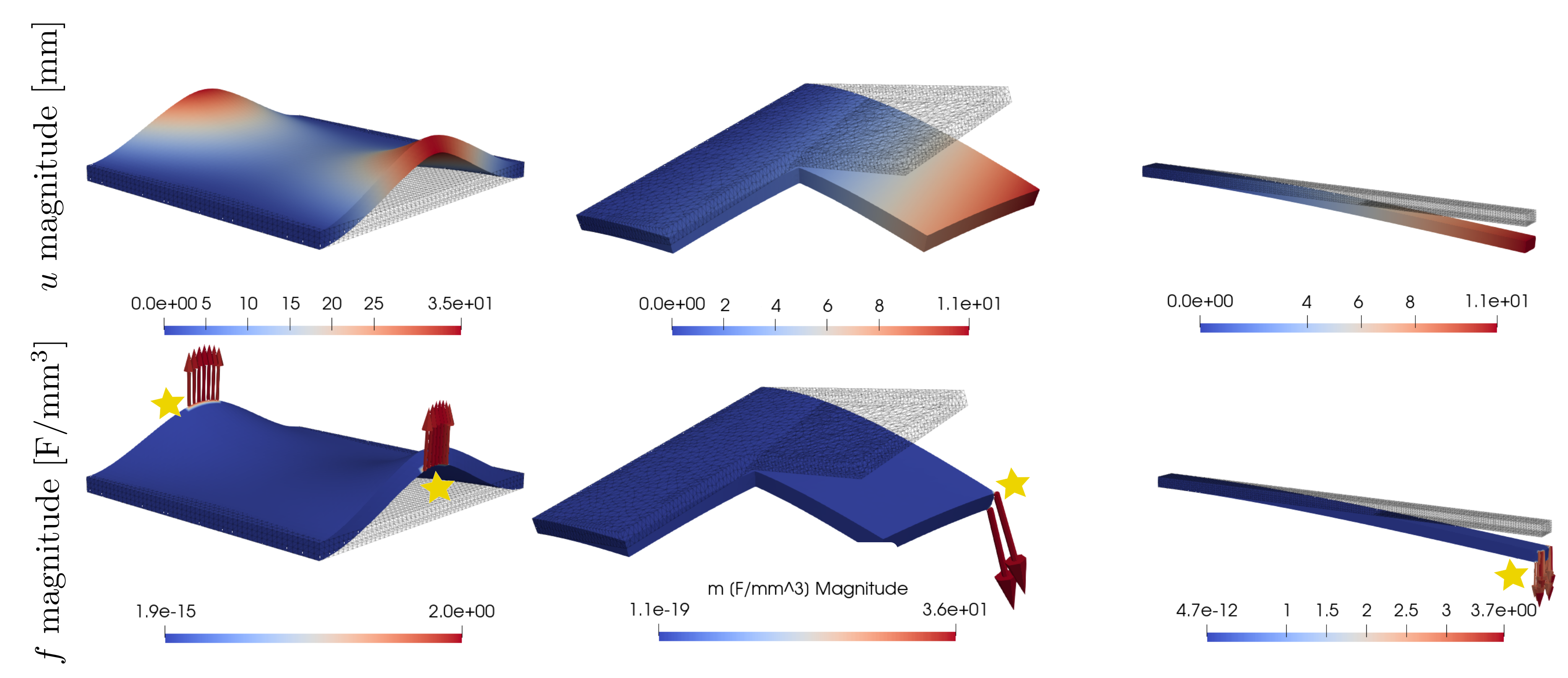} 
	\caption{Identified locations (yellow stars) of minimal stiffness and its direction (glyphs) by means of volume force $\mathbfrm{f}$ and corresponding displacement $\mathbfrm{u}$.}
	\label{fig:minimumstiffnessoptim}
\end{figure}

\begin{figure}[!bht]
	\centering
	\includegraphics[width=140mm]{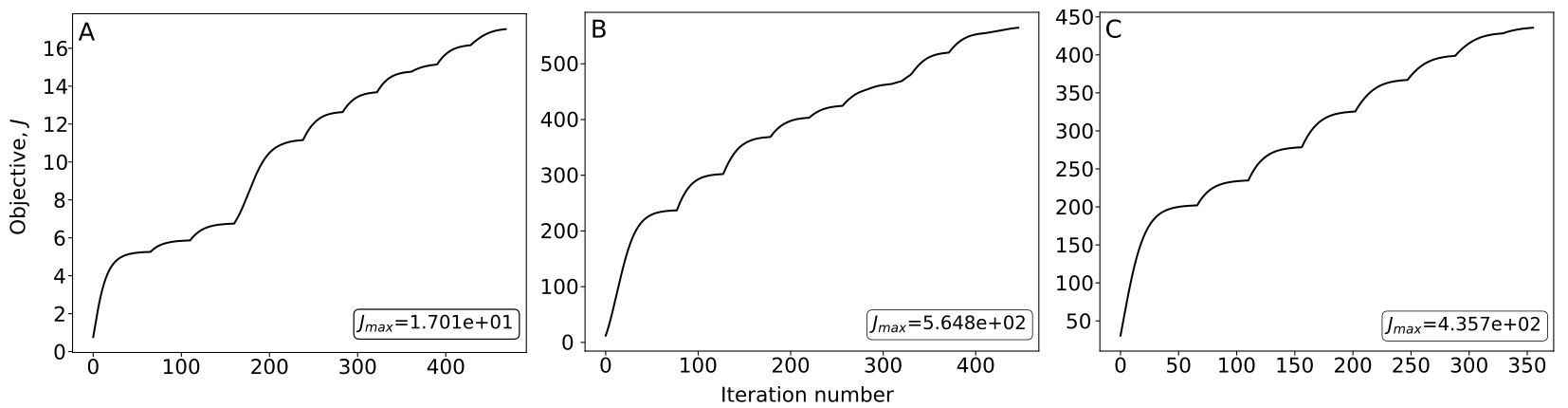}
	\caption{The evolution of the objective functional $\myscalar{J}$ for all three models. The jumps are given by the step increasing of threshold $\myscalar{T}$ in order to keep the gradient of volume force $\myvector{f}$ as smooth as possible.}
	\label{fig:jiterations}
\end{figure}

\begin{figure}[!bht]
	\centering
	\includegraphics[width=1\linewidth]{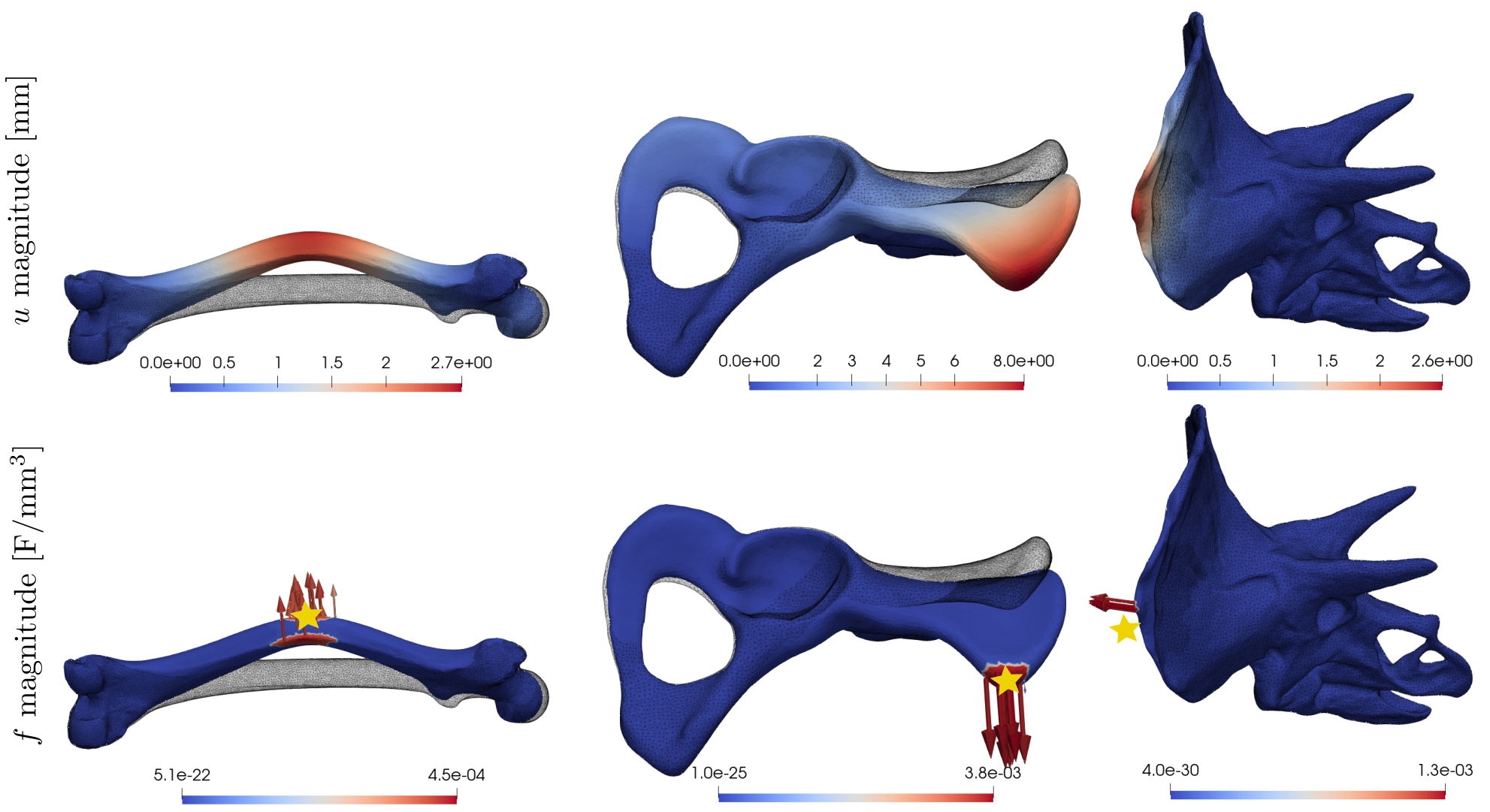} 
	\caption{Identified locations (yellow stars) of minimal stiffness and its direction (glyphs) by means of volume force $\myvector{f}$ and corresponding displacement $\myvector{u}$.}
	\label{fig:complexgeometryoptim}
\end{figure}

\begin{figure}[!bht]
	\centering
	\includegraphics[width=140mm]{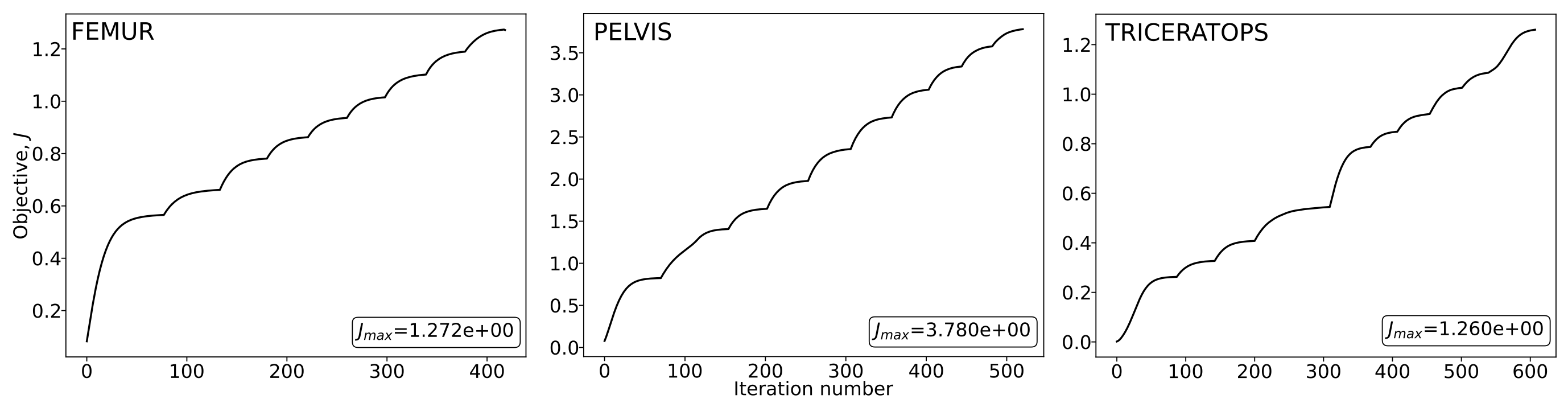}
	\caption{The evolution of the objective functional $\myscalar{J}$ for all three models. The jumps are given by the step increasing of threshold $\myscalar{T}$ in order to keep the gradient of volume force $\mathbfrm{f}$ as smooth as possible.}
	\label{fig:iterationcomplexgeometry}
\end{figure}

In the alternative approach for identifying the critical load configuration based on the numerical optimisation algorithm described in Section \ref{s:gradientbasedmethodtofindcriticalloadingpoint}, the objective functional to be maximised is the deformation strain energy $J \equiv \myscalar{U}=\int_{\Omega}\varepsilon(\myvector{u}):\myvector{\sigma}(\myvector{u})\intdiff{x}$. In Figures \ref{fig:minimumstiffnessoptim} and \ref{fig:complexgeometryoptim}, the location and direction of the normalised force $\myvector{f}^\star$ is shown  together with the corresponding displacement $\myvector{u}$. Visually comparing the results of the spectral analysis in Figures \ref{fig:shapeextremes} and \ref{fig:complegeomshapes} with those obtained by the optimisation approach, we see that the smallest stiffness for the first structure is obtained in the second deformation mode, while the others have the smallest stiffness in the first mode. The PELVIS and FEMUR shapes have the smallest stiffness in the first deformation mode, the TRICERATOPS model -- third deformation mode.
The optimisation algorithm step evaluation is shown in Figures \ref{fig:jiterations} and \ref{fig:iterationcomplexgeometry}. A typical number of iterations was between 300--600, depending on a structure's complexity. The step--like evolution of the objective functional $\myscalar{J} \equiv \myscalar{U}$ (i.e. the elastic strain energy) is given by a threshold stepping described in the subsection \ref{ss:solutionmethod}. The optimal threshold is not given by a single number, but by a sequence of numbers, whose length defines the number of thresholding steps. A typical sequence of thresholding steps was $\{0.1 , 0.27, 0.44, 0.61, 0.78, 0.95\}$. The extreme values (0, 1) are meaningless and destroy the convergence of the algorithm. Moreover, the stepsize $\delta$ defined as a simple rescaling of gradient descent increment $\hat{\myvector{f}}^k$ was typically in a range 0.05--0.2 in order to maintain a stable convergence of the algorithm.
\subsection{Interpretation of Free--free Boundary Conditions}
\label{ss:resultsinterpretationoffreefreeboundaryconditions}
\begin{figure}[!bht]
	\centering
	\includegraphics[width=90mm]{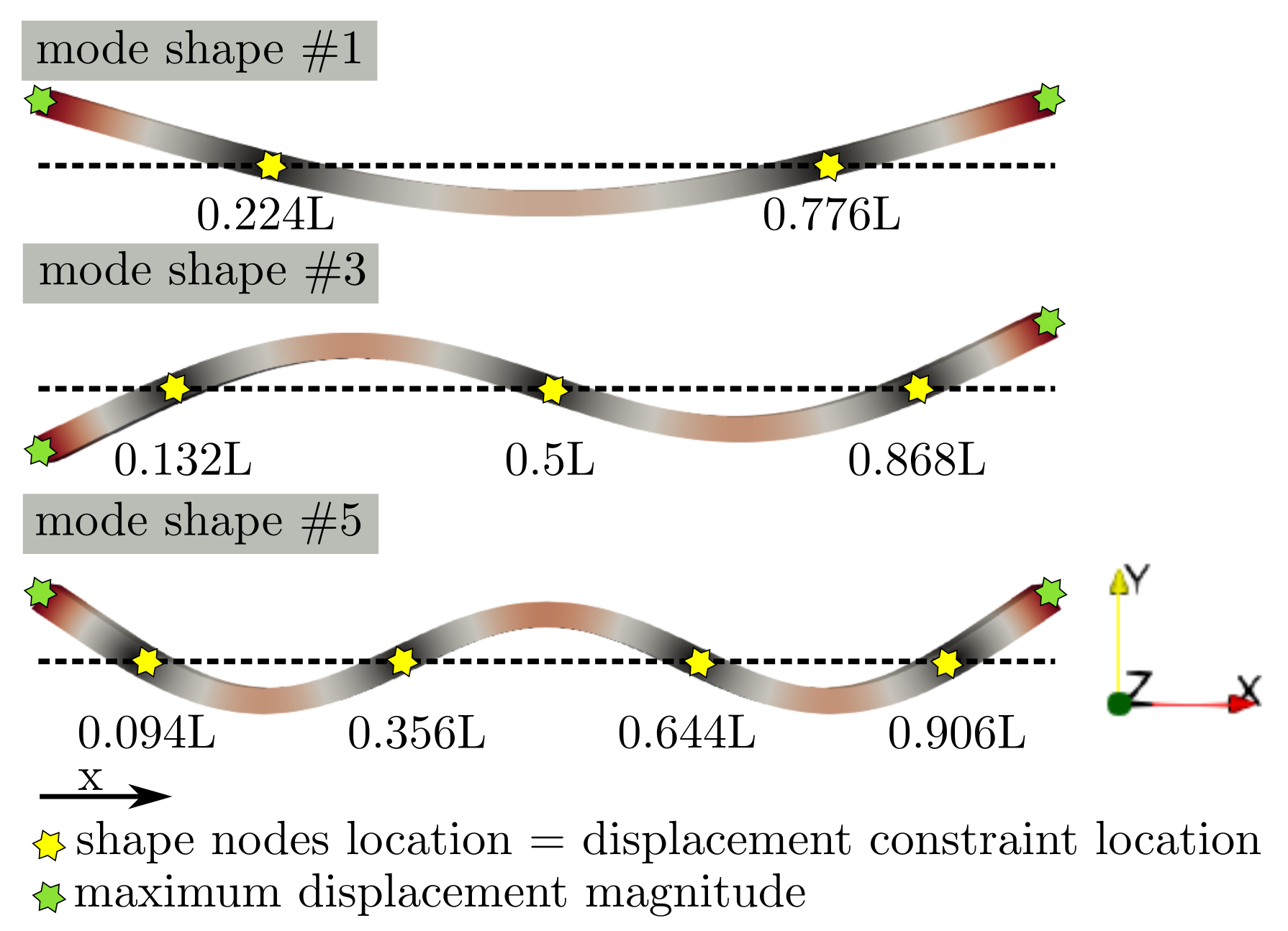}
	\caption{Free bending modal shapes in a plane XY with highlighted minimum displacement (yellow stars).}
	\label{fig:equivalentbcfor-free}
\end{figure}

The free/fixed boundary conditions may be realised by considering alternative pseudo boundary conditions, such that the original free--free shape can be maintained. This is demonstrated in Figure \ref{fig:equivalentbcfor-free}. The free deformation modes contain zero displacement locations (yellow stars), where the alternative point Dirichlet boundary conditions are applied in such a way that the static equilibrium can be maintained. The difference in natural frequencies and modal stiffnesses of a beam model with free and equivalent fixed boundaries is shown in Figure \ref{tab:EigenValues}. Although the values are not exactly the same due to the discretisation, one can see that the smallest absolute difference is 0.1 for the first natural frequency, while the maximal absolute difference is for the modal stiffness represented by the third mode. Note that the similarity of free and equivalent eigenvectors are between 89.5--98.4\% measured by MAC \revised{(Modal assurance criterion)} \cite{allemang2003modal}.
\begin{table}[!bht]
	\caption{A comparison of eigenvalues of free--free beam and equivalent beam bending modes in the plane XY.}
	\label{tab:EigenValues}
	\centering
	\begin{adjustbox}{max width=\textwidth}
		\begin{tabular}{|l|lll|lll|l|}
			\hline
			        & \multicolumn{2}{c}{Natural frequencies [Hz]} & {}         & \multicolumn{2}{c}{Modal stiffness [N/mm]} &            &                                    \\
			Mode \# & Free--Free                                   & Equivalent & difference                                 & Free--Free & Equivalent & difference & MAC [\%] \\
			\hline
			1       & 13.2                                         & 13.1       & 0.1                                        & 17.3       & 17.1       & 0.2        & 98.4     \\
			3       & 36.1                                         & 35.2       & 0.9                                        & 129.1      & 109.6      & 19.5       & 93.6     \\
			5       & 70.1                                         & 67.3       & 2.4                                        & 488.1      & 471.3      & 16.8       & 89.5     \\
			\hline
		\end{tabular}
	\end{adjustbox}
\end{table}
\section*{Discussion}
The spectral analysis of structural stiffness as well as the proposed optimisation algorithm \revised{serve as tools} for analysing the stiffness of a structure from both a quantitative and a qualitative perspective. The stiffness plays an important function in the design of mechanical structures. Its quantification, however, oftentimes requires sophisticated measuring methods. The advantage of the spectral method lies in the fact that the smallest stiffness and its direction can be obtained even for no Dirichlet boundary conditions. However, the absence of Dirichlet boundary conditions poses a difficulty with interpretation of the stiffness as the system is not well defined and can be instead used as a qualitative indicator in order to obtain a general behaviour of a structure itself. In this study, we touch on the question whether the absence of boundary conditions can be modelled by a system with certain boundary conditions such that the resultant spectral properties are equivalent (i.e., the same eigen values and eigen vectors). It was shown that for a simple example of a beam under suitable boundary conditions that mimic the free--free beam supports a sufficiently similar MAC measure and absolute difference of modal/static stiffness and natural frequencies could be obtained. The importance of the free--free conditions is that the system is unaffected by boundary conditions, which can often be imprecise and actually influence the behaviour of the structure itself. Nevertheless, proper analysis and proofs of our suggestion is necessary and will be part of a future study concerning the general stiffness analysis.

The conditions on simple beam examples in Section \ref{ss:beamstiffnessanalysis} for which the $k$'th modal stiffness  approaches the static one from above for two types of boundary conditions have been demonstrated in this study. It was supposed that the $k$'th modal stiffness can approximate the static one when the deformation given by a static load and $k$'th modal deformation associated with modal stiffness were similar. This can be interpreted from the perspective of reduced--order modelling, where only one modal shape is sufficient to approximate the static solution. Nevertheless, for more complex geometries it is not easy to obtain the static load that induces the same deformation as the modal one, see Table \ref{tab:diffsmalleststiffness}. Moreover trying to find a load that induces a modal deformation might not be useful for engineering applications.
\begin{table}[!bht]
	\caption{Relative difference and MAC similarity of smallest static stiffness corresponding deformation shapes from both spectral and optimisation based approaches. The reference for smallest static stiffness and corresponding deformation shape is provided by auxiliary static analysis with point load defined from eigen value analysis. Static stiffness obtained from spectral analysis is approximated by 10 eigen pairs for all tested structures. Low MAC values (bold font values) are given by the fact that the symmetric geometries produce symmetric eigenvectors, which cannot be produced by the static analysis with one point static load.}
	\label{tab:diffsmalleststiffness}
	\centering
	\begin{adjustbox}{max width=\textwidth}
		\begin{tabular}{|l|llllll|}
			\hline
			Model                    & A             & B     & C     & PELVIS & FEMUR & TRICERATOPS   \\
			\hline
			diff [\%] (spectral)     & 9.11          & 2.22  & 0.02  & 6.05   & 10.12 & 14.11         \\
			MAC [-] (spectral)       & \textbf{0.35} & 99.01 & 99.91 & 92.74  & 91.11 & \textbf{0.28} \\
			diff [\%] (optimisation) & 3.57          & 1.98  & 0.01  & 2.89   & 5.79  & 8.54          \\
			MAC [-] (optimisation)   & \textbf{0.35} & 98.41 & 99.1  & 99.1   & 98.1  & \textbf{0.75} \\
			\hline
		\end{tabular}
	\end{adjustbox}
\end{table}
The static stiffness was approximated by a superposition of modal stiffnesses in this study. To obtain a sufficient accuracy of such an approximation, the optimal number of modal contributions should be chosen. For more complex geometrical structures, the distribution of eigenvalues could be more flat and hence for sufficient accuracy a higher number of eigenvalues might be required. Fortunately, most of the structures in this study were simpler and usually up to 10 eigenpairs were needed to get a sufficient accuracy approximation of the static stiffness (see Table \ref{tab:diffsmalleststiffness}). The highest difference of 14.11\% between the static stiffness and the truncated spectral model was observed for TRICERATOPS example. In order to obtain a smaller difference of under 5\%, 20 eigenpairs were required.

The location and direction given by the maximal amplitude eigenvector can still be used to identify the smallest static stiffness of a structure. Nevertheless, due to rescaling of the eigenvalue magnitude, the initial order of the eigenvalues is no longer valid. Although, the matrix $\mymatrix{{\Phi}_\ell}$ can be seen via SVD (after rearrangement of Eq.\eqref{pseudo}), currently the authors are not able to deliver suitable element--wise bounds for its entries (instead of trivial ones) that may improve considerations. Hence the first eigenmode might not point to the smallest static stiffness and the concept of localising the smallest stiffness is rather heuristic.

An alternative to spectral analysis, the adjoint--based optimisation algorithm for finding the maximum compliance (or equivalently smallest stiffness) was developed in this study to compare with the spectral approach. The control field is the volume force field $\myvector{f}$, which is evolved such that the deformation energy is maximised.

Both approaches gave similar results on tested examples. While the spectral approach provides more information about the stiffness of a given structure, its interpretation can be challenging. The difficulties arise from the fact that the magnitude and direction of the smallest static stiffness based on a point compliance tensor $\mytensor{C}$ can be found even for singular problems (i.e., no kinematic boundary conditions are required for modal analysis).

\revised{The assumption of spatially constant Young's modulus is often violated in real mechanical structures and hence the smallest stiffness may be altered. The effect of spatial inhomogeneities in material tensor was not considered in the presented study and it is on a priority list for the next study. A similar effect occurs for cracked structures. Due to the weakening of the structure, the smallest stiffness could dramatically change.}

\section*{Conclusion}
The stiffness of an engineering structure is one of the most important properties whose precise analysis can help design more performant structures or support the decision making process. The authors analysed the structural stiffness using spectral analysis and explored a relationship between the static and modal stiffnesses first analytically for a beam model and then numerically for geometrically more complex structures. Besides the spectral analysis approach, a simple optimisation approach was introduced. Its aim was to find the critical point load configuration that maximised the elastic deformation energy of a given structure. Although both approaches are significantly different, both led to consistent results in terms of point load localisation and its direction. To summarise our findings, the following sub conclusions are made:
\begin{itemize}
	\item Spectral analysis of the stiffness matrix provides useful information about the structure at hand, including localisation and direction of the smallest stiffness (or greatest compliance).
	\item A relation between modal and static compliance/stiffness was explored.
	\item Optimisation algorithm aimed at finding the critical point load configuration for which the structure exhibits the greatest compliance was developed and compared with the spectral analysis approach.
\end{itemize}
\section*{Conflict of interest statement}
The authors declare no conflict of interest.

\bibliographystyle{model1-num-names}
\bibliography{bibliography.bib}

\begin{thebibliography}{27}
\expandafter\ifx\csname natexlab\endcsname\relax\def\natexlab#1{#1}\fi
\providecommand{\bibinfo}[2]{#2}
\ifx\xfnm\relax \def\xfnm[#1]{\unskip,\space#1}\fi
\bibitem[{Zanarini(2019)}]{ZANARINI2019817}
\bibinfo{author}{A.~Zanarini},
\newblock \bibinfo{title}{Full field optical measurements in experimental modal
  analysis and model updating},
\newblock \bibinfo{journal}{Journal of Sound and Vibration}
  \bibinfo{volume}{442} (\bibinfo{year}{2019}) \bibinfo{pages}{817--842}.
\bibitem[{Chen and Maung(2014)}]{CHEN20145566}
\bibinfo{author}{H.-P. Chen}, \bibinfo{author}{T.~S. Maung},
\newblock \bibinfo{title}{Regularised finite element model updating using
  measured incomplete modal data},
\newblock \bibinfo{journal}{Journal of Sound and Vibration}
  \bibinfo{volume}{333} (\bibinfo{year}{2014}) \bibinfo{pages}{5566--5582}.
\bibitem[{Tse(1988)}]{:/content/journals/10.1049/ip-c.1988.0053}
\bibinfo{author}{C.~Tse},
\newblock \bibinfo{title}{Design optimisation of power system stabilisers based
  on modal and eigenvalue-sensitivity analyses},
\newblock \bibinfo{journal}{IEE Proceedings C (Generation, Transmission and
  Distribution)} \bibinfo{volume}{135} (\bibinfo{year}{1988})
  \bibinfo{pages}{406--415(9)}.
\bibitem[{Jureczko et~al.(2005)Jureczko, Pawlak, and Meżyk}]{JURECZKO2005463}
\bibinfo{author}{M.~Jureczko}, \bibinfo{author}{M.~Pawlak},
  \bibinfo{author}{A.~Meżyk},
\newblock \bibinfo{title}{Optimisation of wind turbine blades},
\newblock \bibinfo{journal}{Journal of Materials Processing Technology}
  \bibinfo{volume}{167} (\bibinfo{year}{2005}) \bibinfo{pages}{463--471}.
  \bibinfo{note}{2005 International Forum on the Advances in Materials
  Processing Technology}.
\bibitem[{Wang et~al.(2016)Wang, Kolios, Nishino, Delafin, and
  Bird}]{WANG2016123}
\bibinfo{author}{L.~Wang}, \bibinfo{author}{A.~Kolios},
  \bibinfo{author}{T.~Nishino}, \bibinfo{author}{P.-L. Delafin},
  \bibinfo{author}{T.~Bird},
\newblock \bibinfo{title}{Structural optimisation of vertical-axis wind turbine
  composite blades based on finite element analysis and genetic algorithm},
\newblock \bibinfo{journal}{Composite Structures} \bibinfo{volume}{153}
  (\bibinfo{year}{2016}) \bibinfo{pages}{123--138}.
\bibitem[{Ruggeri et~al.(2010)Ruggeri, Patan{\`e}, Spagnuolo, and
  Saupe}]{ruggeri2010spectral}
\bibinfo{author}{M.~R. Ruggeri}, \bibinfo{author}{G.~Patan{\`e}},
  \bibinfo{author}{M.~Spagnuolo}, \bibinfo{author}{D.~Saupe},
\newblock \bibinfo{title}{Spectral-driven isometry-invariant matching of 3d
  shapes},
\newblock \bibinfo{journal}{International Journal of Computer Vision}
  \bibinfo{volume}{89} (\bibinfo{year}{2010}) \bibinfo{pages}{248--265}.
\bibitem[{Heny{\v{s}} and {\v{C}}apek(2017)}]{henyvs2017material}
\bibinfo{author}{P.~Heny{\v{s}}}, \bibinfo{author}{L.~{\v{C}}apek},
\newblock \bibinfo{title}{Material model of pelvic bone based on modal
  analysis: a study on the composite bone},
\newblock \bibinfo{journal}{Biomechanics and modeling in mechanobiology}
  \bibinfo{volume}{16} (\bibinfo{year}{2017}) \bibinfo{pages}{363--373}.
\bibitem[{Bae et~al.(2019)Bae, You, Suh, and Kang}]{bae2019calculation}
\bibinfo{author}{J.-J. Bae}, \bibinfo{author}{Y.~You}, \bibinfo{author}{J.~B.
  Suh}, \bibinfo{author}{N.~Kang},
\newblock \bibinfo{title}{Calculation of the structural stiffness of run-flat
  and regular tires by considering strain energy},
\newblock \bibinfo{journal}{International Journal of Automotive Technology}
  \bibinfo{volume}{20} (\bibinfo{year}{2019}) \bibinfo{pages}{979--987}.
\bibitem[{Bagheri et~al.(2017)Bagheri, Alipour, Usmani, Ozbulut, and
  Harris}]{bagheri2017structural}
\bibinfo{author}{A.~Bagheri}, \bibinfo{author}{M.~Alipour},
  \bibinfo{author}{S.~Usmani}, \bibinfo{author}{O.~E. Ozbulut},
  \bibinfo{author}{D.~K. Harris},
\newblock \bibinfo{title}{Structural stiffness identification of skewed slab
  bridges with limited information for load rating purpose},
\newblock in: \bibinfo{booktitle}{Dynamics of Civil Structures, Volume 2},
  \bibinfo{publisher}{Springer}, \bibinfo{year}{2017}, pp.
  \bibinfo{pages}{243--249}.
\bibitem[{Feng and Feng(2017)}]{feng2017identification}
\bibinfo{author}{D.~Feng}, \bibinfo{author}{M.~Q. Feng},
\newblock \bibinfo{title}{Identification of structural stiffness and excitation
  forces in time domain using noncontact vision-based displacement
  measurement},
\newblock \bibinfo{journal}{Journal of Sound and Vibration}
  \bibinfo{volume}{406} (\bibinfo{year}{2017}) \bibinfo{pages}{15--28}.
\bibitem[{Jiang et~al.(2018)Jiang, Wu, Ma, and Lin}]{jiang2018structural}
\bibinfo{author}{S.-F. Jiang}, \bibinfo{author}{M.-H. Wu},
  \bibinfo{author}{S.-L. Ma}, \bibinfo{author}{D.-Y. Lin},
\newblock \bibinfo{title}{Structural stiffness identification of traditional
  mortise-tenon joints based on statistical process control chart},
\newblock \bibinfo{journal}{Journal of Aerospace Engineering}
  \bibinfo{volume}{31} (\bibinfo{year}{2018}) \bibinfo{pages}{04018066}.
\bibitem[{MacLeod and Gill(2018)}]{macleod2018exploring}
\bibinfo{author}{A.~MacLeod}, \bibinfo{author}{H.~Gill},
\newblock \bibinfo{title}{Exploring the mechanisms of head-trunnion mechanics
  in modular hip prostheses: The influence of material and structural
  stiffness},
\newblock in: \bibinfo{booktitle}{British Orthopaedic Research Society
  Conference 2018-BORS 2018}.
\bibitem[{Su and Huang(2017)}]{su2017identification}
\bibinfo{author}{W.-C. Su}, \bibinfo{author}{C.-S. Huang},
\newblock \bibinfo{title}{Identification of structural stiffness parameters via
  wavelet packet from seismic response},
\newblock \bibinfo{journal}{Procedia engineering} \bibinfo{volume}{199}
  (\bibinfo{year}{2017}) \bibinfo{pages}{1032--1037}.
\bibitem[{Helsen et~al.(2010)Helsen, Cremers, Mas, and Sas}]{helsen2010global}
\bibinfo{author}{J.~Helsen}, \bibinfo{author}{L.~Cremers},
  \bibinfo{author}{P.~Mas}, \bibinfo{author}{P.~Sas},
\newblock \bibinfo{title}{Global static and dynamic car body stiffness based on
  a single experimental modal analysis test},
\newblock in: \bibinfo{booktitle}{Proceedings of the International Conference
  on Noise and Vibrations Engieneering—ISMA, Leuven, Belgium},
  p.~\bibinfo{pages}{2}.
\bibitem[{Zanardo et~al.(2006)Zanardo, Hao, Xia, and
  Deeks}]{zanardo2006stiffness}
\bibinfo{author}{G.~Zanardo}, \bibinfo{author}{H.~Hao},
  \bibinfo{author}{Y.~Xia}, \bibinfo{author}{A.~J. Deeks},
\newblock \bibinfo{title}{Stiffness assessment through modal analysis of an rc
  slab bridge before and after strengthening},
\newblock \bibinfo{journal}{Journal of Bridge Engineering} \bibinfo{volume}{11}
  (\bibinfo{year}{2006}) \bibinfo{pages}{590--601}.
\bibitem[{Steiger et~al.(2012)Steiger, G{\"u}lzow, Czaderski, Howald, and
  Niemz}]{steiger2012comparison}
\bibinfo{author}{R.~Steiger}, \bibinfo{author}{A.~G{\"u}lzow},
  \bibinfo{author}{C.~Czaderski}, \bibinfo{author}{M.~T. Howald},
  \bibinfo{author}{P.~Niemz},
\newblock \bibinfo{title}{Comparison of bending stiffness of cross-laminated
  solid timber derived by modal analysis of full panels and by bending tests of
  strip-shaped specimens},
\newblock \bibinfo{journal}{European Journal of Wood and Wood Products}
  \bibinfo{volume}{70} (\bibinfo{year}{2012}) \bibinfo{pages}{141--153}.
\bibitem[{West et~al.(2019)West, Capps, Urban, Hartwig, Brown, Bristow,
  Landers, and Kinzel}]{west2019extraction}
\bibinfo{author}{B.~West}, \bibinfo{author}{N.~E. Capps},
  \bibinfo{author}{J.~S. Urban}, \bibinfo{author}{T.~Hartwig},
  \bibinfo{author}{B.~Brown}, \bibinfo{author}{D.~A. Bristow},
  \bibinfo{author}{R.~G. Landers}, \bibinfo{author}{E.~C. Kinzel},
\newblock \bibinfo{title}{Extraction of coupling stiffness of specimens printed
  with selective laser melting using modal analysis},
\newblock in: \bibinfo{booktitle}{Model Validation and Uncertainty
  Quantification, Volume 3}, \bibinfo{publisher}{Springer},
  \bibinfo{year}{2019}, pp. \bibinfo{pages}{275--282}.
\bibitem[{Chandgude et~al.(2020)Chandgude, Gadhave, Taware, and
  Patil}]{chandgude2019investigation}
\bibinfo{author}{N.~Chandgude}, \bibinfo{author}{N.~Gadhave},
  \bibinfo{author}{G.~Taware}, \bibinfo{author}{N.~Patil},
\newblock \bibinfo{title}{Investigation of stiffness of small wind turbine
  blade based on vibration analysis technique},
\newblock \bibinfo{journal}{Wind Engineering} \bibinfo{volume}{44}
  (\bibinfo{year}{2020}) \bibinfo{pages}{49--59}.
\bibitem[{Melnikov et~al.(2017)Melnikov, Soal, and
  Bienert}]{melnikov2017determination}
\bibinfo{author}{A.~Melnikov}, \bibinfo{author}{K.~Soal},
  \bibinfo{author}{J.~Bienert},
\newblock \bibinfo{title}{Determination of static stiffness of mechanical
  structures from operational modal analysis},
\newblock in: \bibinfo{booktitle}{7-th international operational modal analysis
  conference, Ingolstadt, Germany}, pp. \bibinfo{pages}{10--12}.
\bibitem[{Poland et~al.(2015)Poland, Young, Pasha, Allemang, and
  Phillips}]{poland2015estimation}
\bibinfo{author}{J.~Poland}, \bibinfo{author}{A.~Young},
  \bibinfo{author}{H.~Pasha}, \bibinfo{author}{R.~Allemang},
  \bibinfo{author}{A.~Phillips},
\newblock \bibinfo{title}{An estimation of torsional compliance (stiffness)
  from free-free frf measurements: ercf application},
\newblock in: \bibinfo{booktitle}{Experimental Techniques, Rotating Machinery,
  and Acoustics, Volume 8}, \bibinfo{publisher}{Springer},
  \bibinfo{year}{2015}, pp. \bibinfo{pages}{133--139}.
\bibitem[{Deleener et~al.(2010)Deleener, Mas, Cremers, and
  Poland}]{deleener2010extraction}
\bibinfo{author}{J.~Deleener}, \bibinfo{author}{P.~Mas},
  \bibinfo{author}{L.~Cremers}, \bibinfo{author}{J.~Poland},
  \bibinfo{title}{Extraction of static car body stiffness from dynamic
  measurements}, \bibinfo{type}{Technical Report}, SAE Technical Paper,
  \bibinfo{year}{2010}.
\bibitem[{Oliphant(2006)}]{oliphant2006guide}
\bibinfo{author}{T.~E. Oliphant}, \bibinfo{title}{A guide to NumPy},
  volume~\bibinfo{volume}{1}, \bibinfo{publisher}{Trelgol Publishing USA},
  \bibinfo{year}{2006}.
\bibitem[{Wahyuni and Ji(2010)}]{wahyuni2010relationship}
\bibinfo{author}{E.~Wahyuni}, \bibinfo{author}{T.~J.~T. Ji},
\newblock \bibinfo{title}{Relationship between static stiffness and modal
  stiffness of structures},
\newblock \bibinfo{journal}{IPTEK the Journal for Technology and Science}
  \bibinfo{volume}{21} (\bibinfo{year}{2010}).
\bibitem[{Mehmet(2014)}]{avcarM2014}
\bibinfo{author}{A.~Mehmet},
\newblock \bibinfo{title}{Free vibration analysis of beams considering
  different geometric characteristics and boundary conditions},
\newblock \bibinfo{journal}{International Journal of Mechanics and
  Applications} \bibinfo{volume}{4} (\bibinfo{year}{2014}).
\bibitem[{Logg et~al.(2012)Logg, Mardal, and Wells}]{logg2012automated}
\bibinfo{author}{A.~Logg}, \bibinfo{author}{K.-A. Mardal},
  \bibinfo{author}{G.~Wells}, \bibinfo{title}{Automated solution of
  differential equations by the finite element method: The FEniCS book},
  volume~\bibinfo{volume}{84}, \bibinfo{publisher}{Springer Science \& Business
  Media}, \bibinfo{year}{2012}.
\bibitem[{Hu et~al.(2018)Hu, Zhou, Gao, Jacobson, Zorin, and
  Panozzo}]{hu2018tetrahedral}
\bibinfo{author}{Y.~Hu}, \bibinfo{author}{Q.~Zhou}, \bibinfo{author}{X.~Gao},
  \bibinfo{author}{A.~Jacobson}, \bibinfo{author}{D.~Zorin},
  \bibinfo{author}{D.~Panozzo},
\newblock \bibinfo{title}{Tetrahedral meshing in the wild.},
\newblock \bibinfo{journal}{ACM Trans. Graph.} \bibinfo{volume}{37}
  (\bibinfo{year}{2018}) \bibinfo{pages}{60--1}.
\bibitem[{Allemang(2003)}]{allemang2003modal}
\bibinfo{author}{R.~J. Allemang},
\newblock \bibinfo{title}{The modal assurance criterion--twenty years of use
  and abuse},
\newblock \bibinfo{journal}{Sound and vibration} \bibinfo{volume}{37}
  (\bibinfo{year}{2003}) \bibinfo{pages}{14--23}.

\end{thebibliography}

\end{document}